\documentclass[a4paper, amsfonts, amssymb, amsmath, reprint, showkeys, nofootinbib]{revtex4-1}
\usepackage[english]{babel}
\usepackage[utf8]{inputenc}
\usepackage{xcolor}
\usepackage{amsthm}
\usepackage{mathtools}
\usepackage{physics}
\usepackage{xcolor}
\usepackage{graphicx}
\usepackage[left=20mm,right=20mm,top=30mm,columnsep=15pt]{geometry} 
\usepackage{adjustbox}
\usepackage{placeins}
\usepackage[T1]{fontenc}
\usepackage{lipsum}
\usepackage{csquotes}
\usepackage{float}

\usepackage[pdftex, pdftitle={Article}, pdfauthor={Author}]{hyperref} 
\usepackage[switch]{lineno}

\newif\ifproofread
\newcommand{\edit}[1]{%
\ifproofread
\textcolor{red}{#1}%
\else
#1%
\fi
}

\begin{document}
\proofreadfalse
\title{\phantom{paddddd} Individually Addressable \edit{and Spectrally Programmable} \phantom{paddddd} Artificial Atoms in Silicon Photonics}

\author{Mihika Prabhu$^{1,\dagger,*}$, Carlos Errando-Herranz$^{1,2,\dagger,*}$, Lorenzo~De~Santis$^{1,3}$, Ian~Christen$^1$, Changchen Chen$^1$, \edit{Connor Gerlach$^1$,} and Dirk Englund$^{1,}$}
\email{mihika@mit.edu, carloseh@mit.edu, englund@mit.edu}
\address{$^1$Massachusetts Institute of Technology, Cambridge, USA \\$^2$University of M\"unster, M\"unster, Germany \\$^3$QuTech, Delft University of Technology, Delft, Netherlands\\$^\dagger$Equal contribution} 

\date{\today} 

\begin{abstract} 
Artificial atoms in solids have emerged as leading systems for quantum information processing tasks such as quantum networking~\cite{ruf_quantum_2021,bhaskar_experimental_2020,atature_material_2018-1,aharonovich_solid-state_2016-2}, sensing~\cite{degen_quantum_2017}, and computing~\cite{childress_diamond_2013, choi_percolation-based_2019-1}. 
A central goal is to develop platforms for precise and scalable control of individually addressable artificial atoms with efficient optical interfaces. 
Color centers in silicon~\cite{bergeron_silicon-integrated_2020, baron_detection_2021, durand_broad_2021,higginbottom_optical_2022}, such as the recently-isolated carbon-related `G-center'~\cite{hollenbach_engineering_2020, redjem_single_2020}, exhibit emission directly into the telecommunications O-band and can leverage the maturity of silicon-on-insulator (SOI) photonics.
Here, we demonstrate the generation, individual addressing\edit{, and spectral trimming} of G-center artificial atoms in an SOI photonic integrated circuit (PIC) platform. 
Focusing on the neutral charge state emission at 1278~nm, we observe waveguide-coupled single photon emission with an exceptionally narrow inhomogeneous distribution with standard deviation of $1.1$~nm, an excited state lifetime of $8.3\pm 0.7$~ns, and no degradation after \edit{over a month} of operation. 
\edit{In addition, we introduce a technique for optical trimming of spectral transitions up to 300~pm (55~GHz) and local deactivation of single artificial atoms. 
This non-volatile ``spectral programming" enables the alignment of quantum emitters into 25~GHz telecommunication grid channels.}
Our demonstration opens the path to quantum information processing based on implantable artificial atoms in very large scale integrated (VLSI) photonics.
\end{abstract}

\maketitle

\section{Introduction}
Artificial atoms in the solid state are leading candidates for spin-based quantum information processing due to their long spin coherence times and high-quality spin-photon interfaces~\cite{bhaskar_experimental_2020, wan_large-scale_2020-2, wolfowicz_vanadium_2020-1}. 
However, traditional platforms based on diamond and silicon carbide face two critical challenges for large-scale quantum information processing: lack of monolithic manufacturability and inefficient optical interfacing with the optical fiber telecommunication bands. 
Efforts towards alleviating these limitations include hybrid integration~\cite{wan_large-scale_2020-2,kim_hybrid_2020-1,elshaari_hybrid_2020-1} and quantum frequency conversion~\cite{dreau_quantum_2018}, although at the cost of greater optical loss ($>7$~dB at present~\cite{dreau_quantum_2018}).
Recently, a number of color centers in silicon~\cite{baron_detection_2021, redjem_single_2020, bergeron_silicon-integrated_2020, hollenbach_engineering_2020}, such as the carbon-based G-center~\cite{redjem_single_2020, hollenbach_engineering_2020}, have emerged as promising qubit candidates, as they can be integrated natively with existing commercial silicon platforms and their telecom wavelength emission obviates the need for frequency conversion.  
Furthermore, the lack of a nuclear spin bath in isotopically purified $^{28}$Si has allowed demonstrations of electron and nuclear spin coherence $T_2$ for donors in silicon exceeding 2~seconds~\cite{tyryshkin_electron_2012} and 39~min~\cite{saeedi_room-temperature_2013-1} respectively.

\begin{figure*}[thbp]
  \centering
  \includegraphics[width=\textwidth]{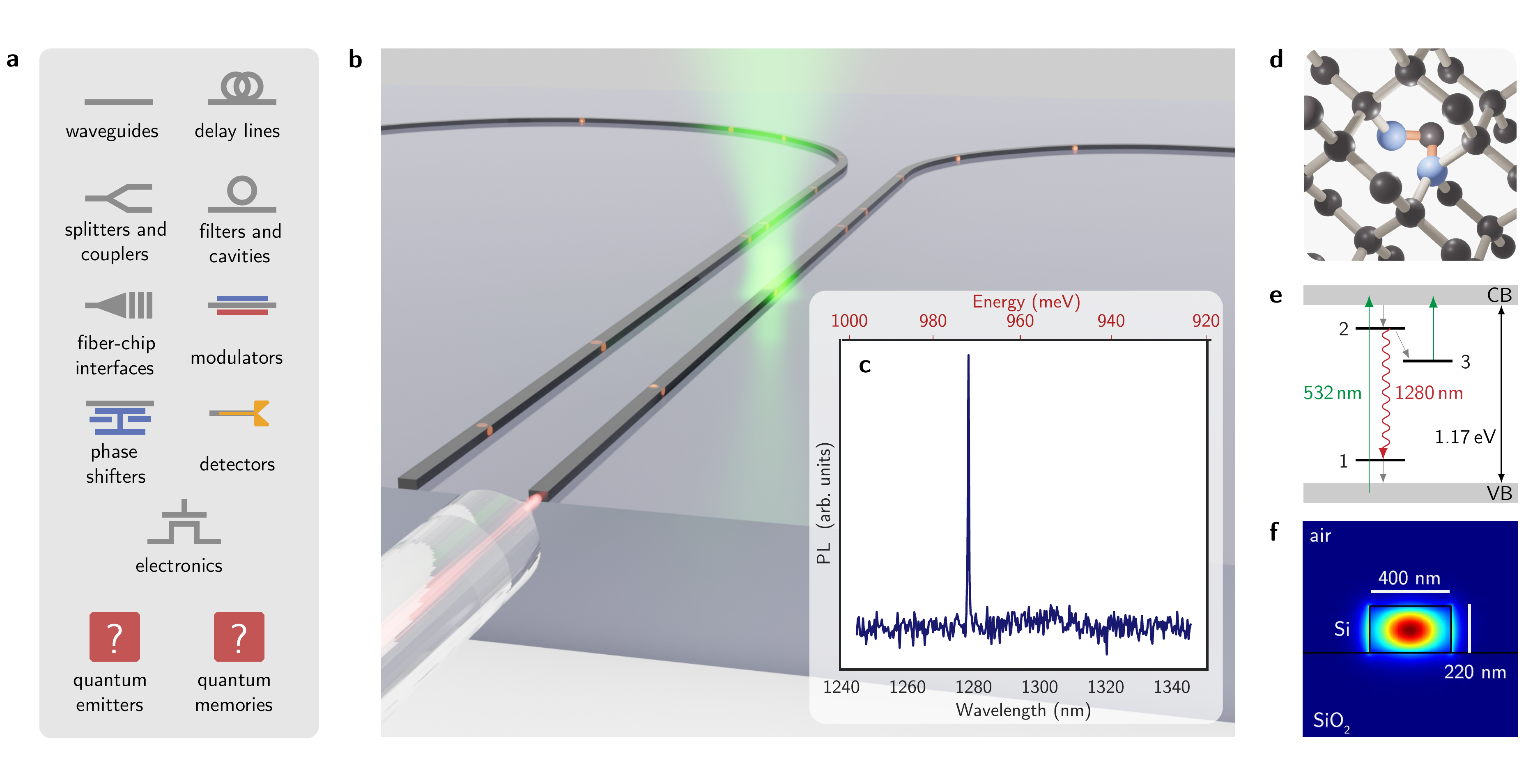}
\caption{\textbf{G-centers in silicon photonics as a scalable quantum photonic platform.} a) The silicon platform features technologically mature opto-electronics, but lacks integrated quantum emitters and memories. b) Schematic showing our system, which integrates artificial atoms (G-centers) in silicon photonic waveguides. c) Measured G-center PL spectrum through our waveguide, showing light emission in the O-band. d) The G-center atomic structure consists of two substitutional carbon atoms (blue) bonded to an interstitial silicon and e) has two optically addressable energy levels and a third metastable level within the silicon bandgap. f) A finite-difference eigenmode simulation showing the electric field norm for our single-mode waveguide geometry.}
  \label{fig1}
\end{figure*}

An artificial atom platform based on the silicon G-center could immediately access a vast manufacturing and science toolkit, including the world's most advanced complementary metal-oxide-semiconductor (CMOS) platforms, which already include carbon defects~\cite{plummer_silicon_2000} and large-scale patterning with nanometer-scale resolution~\cite{Yeap_5nm_2019-1}.
Moreover, VLSI silicon photonics can integrate millions of devices onto a single wafer, and already includes nearly all the necessary components for a full-stack quantum photonic system: low-loss passive components such as waveguides, splitters, fiber-to-chip interfaces~\cite{vivien_handbook_2016}, and  ultra-high quality factor cavities~\cite{panuski_fundamental_2020}; high-speed and cryogenically-compatible active modulators~\cite{chakraborty_cryogenic_2020} and phase shifters~\cite{edinger_silicon_2021, gyger_reconfigurable_2021}; superconducting single-photon detectors~\cite{pernice_high-speed_2012-3}, and control electronics~\cite{giewont_300-mm_2019-1} (Fig.~\ref{fig1}a).

Although dense ensembles of silicon artificial atoms have been recently integrated into photonic waveguides~\cite{tait_microring_2020, buckley_optimization_2020}, the isolation of single silicon artificial atoms in PICs remains a central challenge.
\edit{While the integration of quantum emitters in PICs in other material systems has already resulted in a plethora of new physical breakthroughs and applications~\cite{lodahl_interfacing_2015}, in silicon,} a platform that combines individually-addressable G-centers with mature integrated photonics could open the door to industrial-scale quantum photonics.
Such a technology has the potential to address the scalability challenges of quantum information processing.

Here we report the first monolithic integration of single artificial atoms -- G-centers -- in silicon PICs (Fig.~\ref{fig1}b) \edit{and their optically-induced non-volatile spectral programming}. 
Low temperature spectroscopy reveals single photon emission in the telecommunications O-band (Fig.~\ref{fig1}c) with a narrow inhomogeneous distribution of $1.1$~nm \edit{and spectral shifts up to 300~pm (55~GHz)}. 

The silicon G-center consists of a pair of substitutional carbon atoms bonded to a silicon self-interstitial (C$_\text{s}$-Si$_\text{i}$-C$_\text{s}$) within the silicon crystal lattice (see Fig~\ref{fig1}d). 
It exhibits a zero phonon line (ZPL) transition at 970~meV (1278~nm), resulting from an electron transition between $sp$-like orbitals localized at the Si$_\text{i}$ atom (Fig~\ref{fig1}e) and features a spin triplet metastable state~\cite{udvarhelyi_identification_2021}. 
While the primary mechanism for above-band population of the excited state in G-centers is likely to be Shockley-Read-Hall (SRH) recombination, additional non-radiative mechanisms such as surface recombination at electronically active surface states and Auger recombination are also known to affect carrier dynamics in silicon~\cite{coldren_diode_2012} (a more detailed discussion can be found in SI Section~\ref{sec:rates}).

\begin{figure*}[hbtp]
  \centering
  \includegraphics[width=\textwidth]{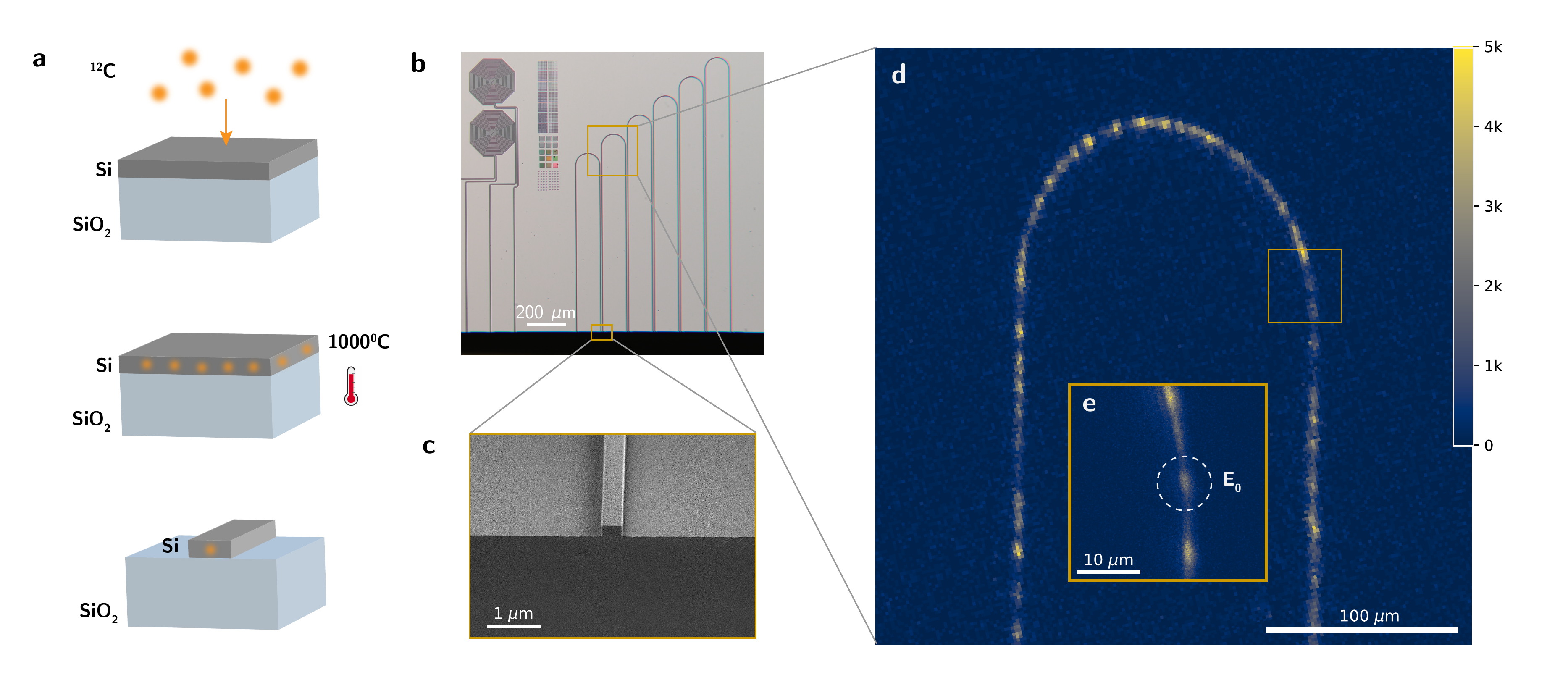}
\caption{\textbf{Fabrication and imaging of G-centers in silicon photonics.} a) Fabrication process for G-centers in silicon photonic waveguides. b) Microscope image of our sample, showing several silicon photonic structures. c) Scanning electron micrograph (SEM) of a cleaved facet of one of our waveguides. d) A PL map taken under 10$\mu$W 532~nm green continuous-wave excitation shows discrete light emitters coupled into the waveguide. e) Emitter~0 under study.}
  \label{fig2}
\end{figure*}

The device under study consists of an SOI waveguide designed for single-mode operation at 1278~nm, the ZPL of the G-center. 
Simulation results for the electric field distribution within the waveguide are shown in Figure~\ref{fig1}f.
We fabricated our samples by a combination of commercial carbon implantation, thermal annealing, and foundry electron beam lithography and etching (see Fig.~\ref{fig2}a and Methods). 
Our sample contains several waveguides and photonic structures.
The waveguides in our sample (see Fig.~\ref{fig2}b) start and end on a cleaved facet (Fig.~\ref{fig2}c) and loop in a $63.5$~\textmu m radius bend.


We characterized the samples using low-temperature spectroscopy in a cryostat at a base temperature of $6$~K. 
Using above band-gap laser excitation (wavelength $\lambda_\text{exc}=532$~nm, NA=0.55, and power $P_\text{in}$ measured before the objective), we induced artificial atoms to emit into the waveguide.
The waveguide-coupled optical emission from the emitters was subsequently collected with an edge-coupled lensed fiber at the cleaved chip facet. 
We band-pass filtered the fiber-collected emission between $1250$~nm and $1550$~nm to isolate the zero-phonon line (ZPL) and phonon sideband (PSB) of the G-centers while attenuating residual pump light and other background. 
The collected emission was then detected with either a cryogenic superconducting-nanowire single-photon detector (SNSPD) system or an infrared spectrometer.

We spatially locate our artificial atoms within the waveguides using photoluminescence (PL) raster scans.
We scanned the focused continuous-wave green excitation along the chip plane and detected the filtered emission through the fiber-coupled waveguide with an electronically-gated SNSPD. 
Figure~\ref{fig2}d shows PL signal in spatially-isolated loci along the waveguide, corresponding to single emitters or small ensembles of emitters.
A range in brightness of the PL hotspots can be attributed to the presence of clusters of emitters in close proximity, which is confirmed through the observation of multiple distinct ZPL peaks in the PL spectra taken from these points (SI Section~\ref{sec:background} and Fig.~\ref{fig:backgroundsubtraction}b). 

\begin{figure*}[htbp]
  \centering
  \includegraphics[width=\textwidth]{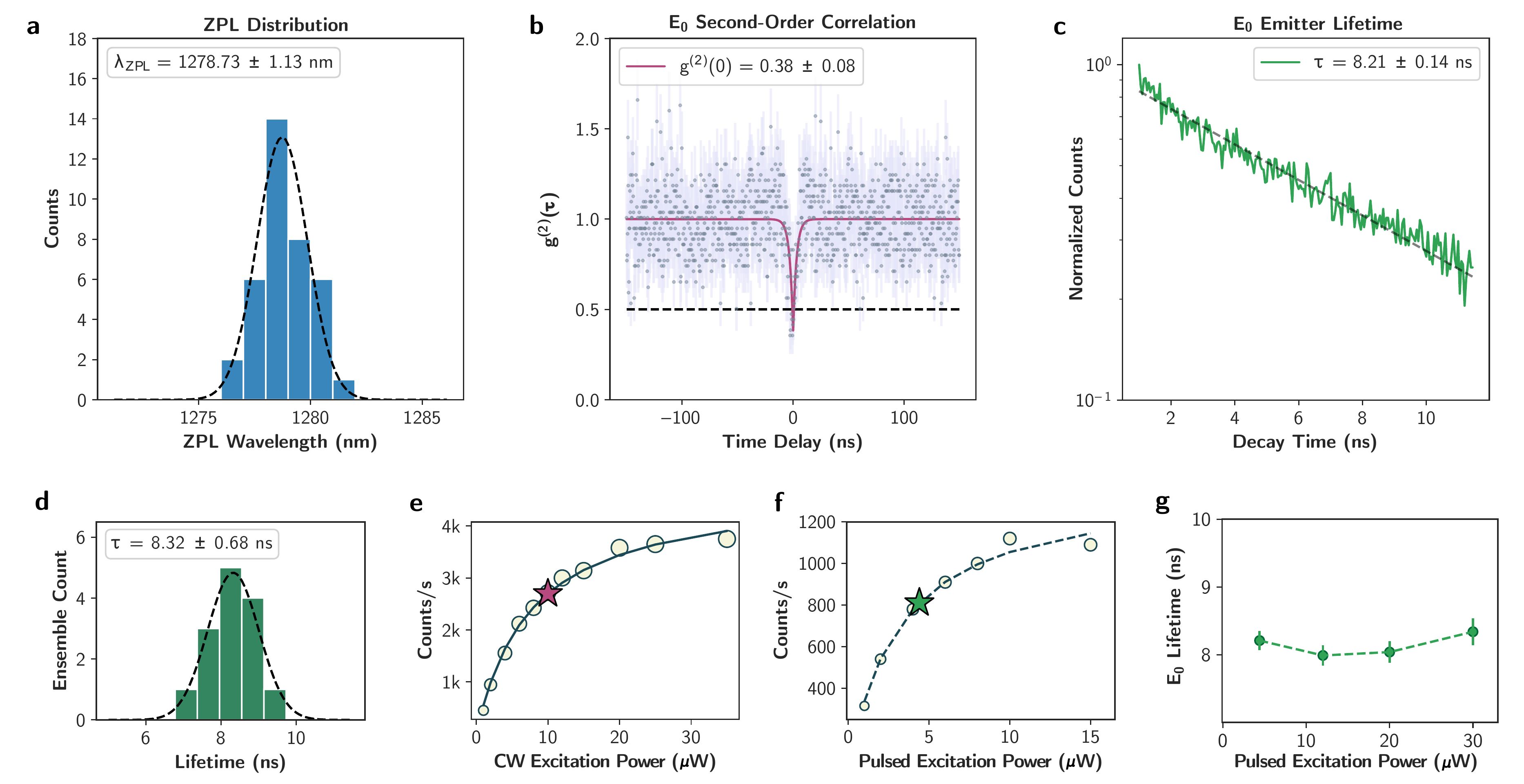}
\caption{\textbf{Isolation of single emitters and statistics.} a) ZPL distribution of 37 waveguide-coupled G-centers. b) Second-order autocorrelation measurement for E$_0$ shows g$^{(2)}(0)<0.5$ and single photon emission \edit{for a continuous-wave excitation power of 10~\textmu W, corresponding to an estimated power density of $5.4$~kW$\cdot\text{cm}^{-2}$}. c) Lifetime E$_0$ and d) statistics for 13 G-centres. Saturation curves under e) continuous-wave and f) pulsed excitation fit well to a two-level system. Error bars are denoted by the radii of the circle markers. \edit{The excitation power used for measurements b) and c) are marked with purple and green stars respectively.} g) The measured lifetime remains constant under an order of magnitude variation in excitation power.}
  \label{fig3}
\end{figure*}

Figure~\ref{fig3}a shows that the ZPL distribution of 37 waveguide-coupled G-centers matches a Gaussian probability distribution with a standard deviation of $\sigma_\text{inh}=1.1$~nm, nearly an order of magnitude narrower than previous reports~\cite{redjem_single_2020}.
Our narrower inhomogeneous distribution may be due to strain relaxation induced by our waveguide geometry (see SI Section~\ref{sec:inhomstrain}).
The fitted mean ZPL wavelength is $1278.7$~nm, in agreement with prior results in bulk SOI~\cite{redjem_single_2020, beaufils_optical_2018}. 
Using the number and width of the ZPL peaks in the PL spectrum as an indication of the number of individual emitters within each excitation region, we identified locations within the waveguide that are likely to contain single emitters. 
We isolated one such area, denoted ``Emitter 0'' (E$_{0}$, shown in Fig.~\ref{fig2}e), and characterized the photophysics of the artificial atom in this region. 
The PL spectrum from E$_{0}$ is shown in Fig.~\ref{fig1}e and exhibits a single resolution-limited peak at $1278.473\pm0.155$~nm. 
To confirm the presence of a single artificial atom at E$_0$, we performed second-order autocorrelation (Fig.~\ref{fig3}b) and power-dependent emission intensity measurements (Fig.~\ref{fig3}c).

Focusing continuous-wave excitation at 532~nm on E$_0$, we measured the dependence of the emission count rate on the excitation power. 
In order to distinguish the contribution to the count rate generated by the emitter from the linear fluorescence observed in the waveguide, we performed background subtraction of the count rate (see Methods and SI Section~\ref{sec:background}). 
The background-corrected emission fits well to the characteristic two-level emitter saturation model~\cite{beaufils_optical_2018}, given by 
\begin{eqnarray}
    I(P)=I_{\infty}\frac{P}{P+P_\text{sat}},
\end{eqnarray}
with a saturation power $P_{\text{sat}}^{\text{cw}}=7.6\pm0.5$~\textmu W, and a saturation intensity \edit{under continuous wave (CW) laser excitation of} $I_{\infty}^{\text{cw}}=4753\pm122$ counts per second (Fig.~\ref{fig3}e). 
\edit{The extracted saturation power corresponds to an estimated continuous-wave power density of $4.1$~kW$\cdot\text{cm}^{-2}$, assuming a diffraction-limited spot size with objective NA=0.55. 
We note, however, that the true excitation area is bounded in one axis by the sub-diffraction waveguide width of 400~nm.}
These results closely match the saturation power observed for single isolated G-centers in unpatterned SOI wafers~\cite{redjem_single_2020}.
We performed a similar background-subtracted measurement for 532~nm pulsed excitation (see Methods and SI Section~\ref{sec:lifetime} for extended details), resulting in a pulsed saturation average power of $P_{\text{sat}}^{\text{pulsed}}=3.1\pm0.4$~\textmu W (Fig.~\ref{fig3}f). 

Using continuous-wave excitation power of $10$~\textmu W (slightly above saturation), we performed second-order autocorrelation measurements using Hanbury-Brown-Twiss interferometry. 
We split the resulting fiber-collected emission from our sample with a 50:50 O-band fiber beam splitter, followed by detection with two time-tagged SNSPDs (detection efficiencies of 21\% and 24\% at $1280$~nm). 
We fitted the measured histogram of coincidences as a function of time delay between the detection events to the second-order autocorrelation function of a two-level system emitter, displaying an antibunching dip of g$^{(2)}(0)=0.38\pm0.08$ (Fig.~\ref{fig3}b, see Methods for additional details). 
The observed antibunching dip with g$^{(2)}(0)<0.5$ confirms single-photon emission and individual addressing of a single artificial atom coupled to a silicon photonic waveguide.

We subsequently measured the excited-state lifetime statistics of emitter E$_0$ and 13 other G-center ensembles (a total of 14 spots) using pulsed 532~nm excitation power of $4.4$~\textmu W, slightly above the measured saturation power for a single emitter.  
Resulting decay curves fit well to a mono-exponential function, following clipping to remove laser leakage and background (see Methods and SI Section~\ref{sec:lifetime}). 
The lifetime distribution of the 14 measured G-center ensembles fits well to a Gaussian distribution with mean lifetime $8.33$~ns and a standard deviation of $0.68$~ns (Fig.~\ref{fig3}d).
The lifetime of the single G-center at E$_0$ was measured to be $8.21\pm0.14$~ns (Fig.~\ref{fig3}c), in close agreement with the mean lifetimes of the measured ensembles. 
Additionally, we measured the lifetime of the single emitter E$_0$ to be constant over an order of magnitude variation in excitation power (Fig.~\ref{fig3}g). 
Our results agree with previously reported lifetimes in bulk SOI G-center ensembles~\cite{beaufils_optical_2018}, but indicate shorter lifetimes compared to prior results on isolated G-centers in unpatterned SOI wafers~\cite{redjem_single_2020}.
Calculations of the dipole local density of optical states (LDOS) in our waveguide were compared to the LDOS of a dipole emitter in a SOI slab (see SI Section~\ref{sec:coupling}) and suggest that the reduced lifetime we observe is not explained by field enhancement in the patterning alone. 
Differences in measured lifetimes in our waveguides could be attributed to increased surface recombination in etched surfaces (see SI Section~\ref{eqn:rateequations}).
Other effects affecting the measured lifetimes may be doping and defect density variations between wafers. 
A quantitative characterization of these effects would include geometry-dependent lifetime measurements, which are out of the scope of this study. 

\begin{figure*}[htbp]
  \centering
  \includegraphics[width=\textwidth]{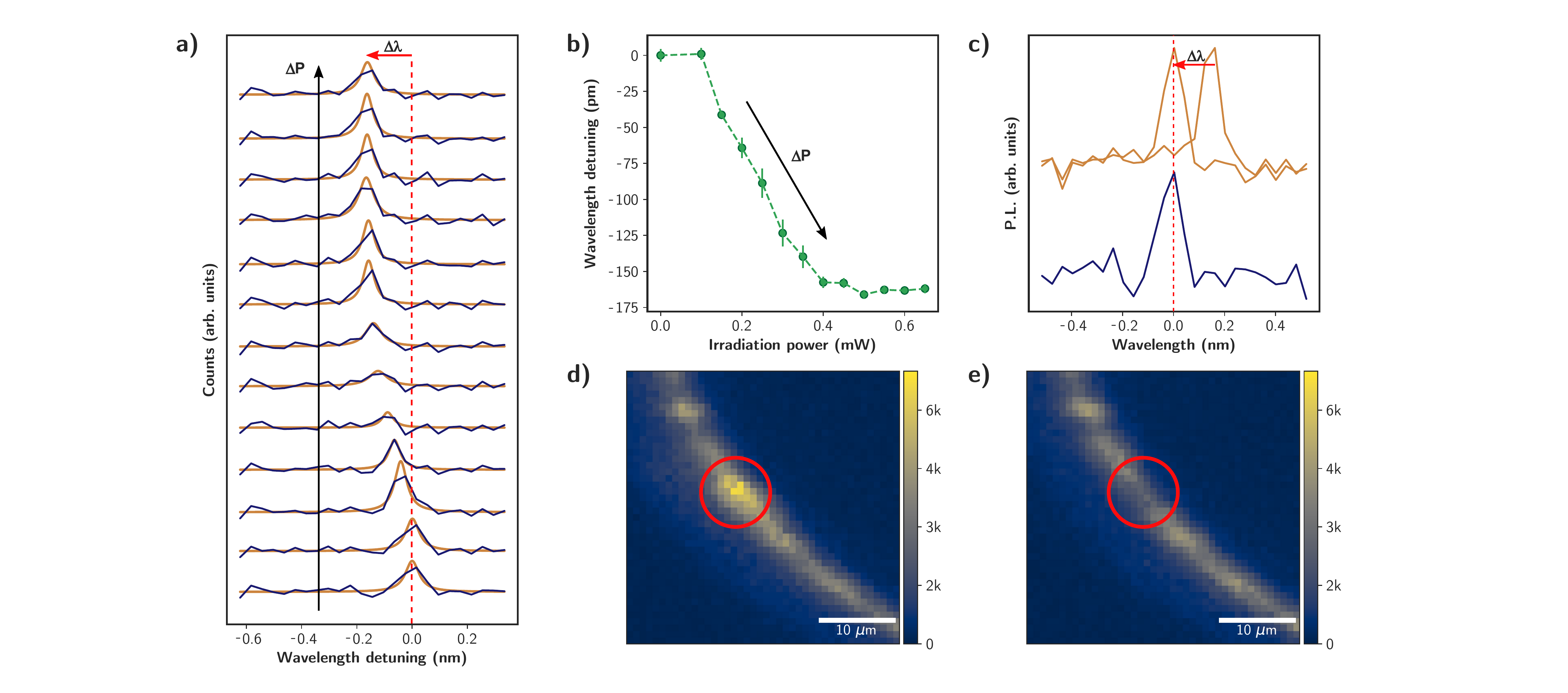}
\caption{\edit{\textbf{Non-volatile trimming of color centers using light.} a) Example of non-volatile spectral shift of the ZPL of artificial atoms caused by optical irradiation. b) Fitted ZPL central wavelength under increasing optical irradiation power. c) Our trimming technique enables non-volatile spectral alignment of separate silicon color centers. d) and e) show PL maps of waveguide sections before (d) and after (e) local deactivation of an emitter (marked with a red circle).}}
  \label{fig:tuning}
\end{figure*}

\edit{
Moreover, we demonstrate non-volatile spectral trimming and deactivation of single color centers in PICs.
Our ``spectral programming'' process consists of in-situ local irradiation of G-centers with a 532~nm CW laser with power above 0.1~mW (estimated power density of $54.4$~kW$\cdot\text{cm}^{-2}$) during 15~s in our 6~K cryostat, followed by probing PL measurements near emitter saturation powers.
Under moderate irradiation powers above 0.1~mW, we observe consistent non-volatile spectral shifts of the ZPL for 11 out of our 12 probed emitters. 
We observe controllable shifts of average amplitude 150~pm (27.5~GHz) and up to 300~pm (55~GHz) as shown in Fig.~\ref{fig:tuning}a and b and Fig.~\ref{fig:tuningall}), large enough to match the 25~GHz telecommunication bands and to enable spectral alignment of separate emitters (Fig.~\ref{fig:tuning}c).
Figure~\ref{fig:tuningstab}a and \ref{fig:tuningstab}b show the spectral stability of our programmed emitters.
Higher powers in the order of 1~mW (estimated power density of $544.3$~kW$\cdot\text{cm}^{-2}$) result in broadening and deactivation of the emitter, leaving the waveguide and adjacent emitters unaffected (Fig.~\ref{fig:tuning}d and e and the subtracted PL map in Fig.~\ref{fig:deactivation}).
In our experiments, the programming and deactivation effects are non-reversible.
Given that our simulations rule out local annealing as a cause for the trimming and deactivation effects (see SI Section~\ref{sec:tuning}), we hypothesize that these effects are caused by optically-induced variation of surface charges leading to Stark tuning of the emitters, followed by ionization into the dark A state~\cite{udvarhelyi_identification_2021} for high charge densities.
For more information on the process, results, and our hypotheses on the physics behind this effect, see SI Section~\ref{sec:tuning}.
}

The count rates in our waveguide measurements were limited by preventable loss due to imperfect mode matching between our cleaved waveguide facet and the lensed collection fiber. 
A mode overlap simulation between the SOI waveguide and fiber mode predicts an upper bound on the coupling efficiency in our system to be $8.25$\% (see SI Section~\ref{sec:coupling}). 
In practice, this excess loss restricts our excitation to a small range near the saturation power of the emitter, as the signal-to-noise ratio of the emission decreases at low excitation powers due to the detector dark counts and also at high excitation powers due to the background fluorescence (see SI Section~\ref{sec:background}). 
This low signal to noise ratio is the reason behind our non-zero g$^{(2)}(0)$.
However, the coupling efficiency can be easily improved by up to an order of magnitude using widely available photonic components such as spot-size converters or grating couplers, that can achieve coupling efficiencies in excess of $80$\%~\cite{zhang_low-cost_2020, chen_dual-etch_2017}.

Another notable difference between our waveguide-integrated G-centers compared to those measured in unpatterned SOI wafers concerns the position of observed emitters selected by the collection mode.
Solid-state emitters often show decreased spectral stability in proximity to surfaces due to emitter-surface interactions~\cite{chakravarthi_impact_2021}.
Simulations of dipole collection efficiency as a function of dipole depth within our waveguides suggest that we preferentially observe emission from G-centers near the center of the silicon film (SI Section~\ref{sec:coupling}, Fig.~\ref{fig:effsims}b). 
On the other hand, prior results using confocal collection of G-center emission from silicon slabs select for emitters close to the surface~\cite{redjem_single_2020}. 
\edit{This geometrical filtering effect may also explain the absence in our measurements of other emitters previously observed in silicon~\cite{durand_broad_2021}, or those created by etching processes.}
Our measurements show high emission stability down to a 10~ms timescale for a range of excitation powers\edit{, as well as robustness to photobleaching, enabling emission measurements over time intervals greater than one month} (see SI Section~\ref{sec:stab}).
A comparative study looking into G-center emission at varying depths would required to probe the emitter-surface interaction for these artificial atoms, but this falls out of the scope of this work.

\edit{
Our in-situ non-volatile trimming results show a path towards post-fabrication fine tuning of artificial atoms.
Such an effect is large enough to tune emitters into standard 25~GHz telecommunication bands, to align emitters to cavities and enhance light-matter interaction, or to spectrally align separate emitters and achieve quantum interference or cooperative emission. 
In addition, the local deactivation of emitters presented here can aid in trimming waveguides and cavities of unwanted artificial atoms.
Further work is required to reveal the physics behind the observed effect and potentially increase its range.
This may involve performing higher resolution spectroscopy, measuring samples with different geometries and crystal purity, experimenting with pulsed lasers and other wavelengths, performing in-situ temperature measurements, measuring its dynamics, or searching for correlations between trimming performance and dipole orientation.
}

Efficient integration with PICs is a key requirement for any large-scale artificial atom qubit platform.
Therefore, the results presented here characterizing single G-center emission \edit{and trimming} in silicon-on-insulator waveguides provide a key step forward for quantum information processing based on color centers in silicon. 
\edit{Moreover, our use of a commercial foundry for the fabrication process makes our results scalable and inherently repeatable by the scientific community, without the need to re-develop an in-house fabrication process.}
Promising future research directions include \edit{the} Purcell-enhanced emission from single G-centers in resonant structures and \edit{the} generation of spectrally indistinguishable single photons for quantum interference \edit{using optical trimming,} electric fields~\cite{anderson_electrical_2019-2}, or \edit{mechanical} strain~\cite{wan_large-scale_2020-2}. 
\edit{As demonstrated in other color centers~\cite{bhaskar_experimental_2020}, coupling emitters into cavities help overcome low quantum efficiencies and Debye-Waller factors.
This may not only aid in the development of spin-optic interfaces, but also may enable deterministic single-photon emission and photon-photon interactions via cavity quantum electrodynamics~\cite{janitz_cavity_2020}.}
Investigation of the spin properties of the G-center metastable state~\cite{udvarhelyi_identification_2021}\edit{, particularly its spin lifetime,} and the $^{13}$C and $^{29}$Si nuclear degrees of freedom~\cite{wolfowicz_29si_2016} could additionally enable optically-active quantum memories hosted in silicon photonics. 
Finally, our results also motivate the study and waveguide integration of other radiation damage centers in silicon, such as the recently isolated T-center~\cite{bergeron_silicon-integrated_2020} and W-center~\cite{baron_detection_2021}.

In conclusion, we demonstrated individually addressable artificial atoms operating at telecommunication wavelengths and featuring narrow inhomogeneous distributions in a foundry-written silicon photonic circuit\edit{, as well as a new method to spectrally program them and deactivate them using light}.
Our results show native \edit{and spectrally programmable} single-photon emission and pave the way towards spin qubits in silicon waveguides, establishing silicon photonics as a \edit{promising} platform for large-scale quantum information technologies.
\vspace{-5pt}

\section{Methods}
\subsection{Sample description}
The device under study consists of a background p-doped silicon-on-insulator waveguide with $2$~\textmu m silicon dioxide bottom cladding on a silicon substrate. 
The cross-sectional geometry is shown in Fig.~\ref{fig1}f and is rectangular with $400$~nm width and $220$~nm height. 
The waveguide starts and ends on a cleaved chip facet and loops in a $63.5$~\textmu m radius $180^{\circ}$ bend.

\subsection{Sample fabrication}
We generated silicon G-centers using a fabrication process that follows~\cite{redjem_single_2020}. 
The samples started from a commercial (SOITEC) unclad silicon-on-insulator wafer ($220$~nm Si on $2000$~nm SiO$_2$). 
The wafer was then cleaved into $1$~cm$^2$ pieces, implanted with $^{12}$C with a dose of $1\times10^{13}$~ions/cm$^{-2}$ at $36$~keV energy and at an angle of $7^\circ$, and then flash annealed for $20$~s at $1000~^{\circ}$C.
The sample was then electron-beam patterned and etched in a foundry (Applied Nano Tools).
\edit{The silicon etching was performed using inductively coupled plasma reactive ion etching with SF$_6$-C$_4$F$_8$ mixed-gas, in a process optimized for vertical sidewall etching and low propagation loss.}
This process resulted in silicon waveguides with SiO$_2$ bottom cladding and air as top cladding.
To enable fiber coupling, the sample was cleaved across the waveguides.

\subsection{Photophysics characterization}
Optical excitation of the G-centers was performed through the cryostat window using long working distance objectives with numerical aperture of $0.55$ .
Waveguide-coupled emission was collected from a single waveguide output using a SMF-28 lensed fiber and spectrally filtered in free space to select signal between $1250$ and $1550$~nm. 
The signal \edit{was} then detected using either time-tagged SNSPD or an infrared spectrometer with a wavelength resolution of 155~pm. 
We measured the excitation powers for all presented measurements immediately prior to the microscope objective. 

Photoluminescence (PL) raster scans were acquired by scanning the focused excitation spot over a spatial region of the waveguide sample and gating the SNSPD integration with the electronic trigger of the scanning mirrors. 
Background-corrected PL spectra were then measured using the infrared spectrometer, where the background spectrum recorded environmental light conditions with the laser blocked. 

Saturation curves were measured at a spatial location in the waveguide that exhibited bright PL intensity and a ZPL peak near the 1280~nm transition of the G-center. 
Correction of the count rate due to \edit{the} waveguide background was performed prior to fitting with the saturation model in Eq. (1) (see SI Section~\ref{sec:background} for details on background measurements and subtraction).

The second-order correlation (g$^{(2)}(\tau)$) was measured with continuous-wave 532~nm excitation near the saturation power. 
Emission from the waveguide was split with a 50:50 O-band fiber beamsplitter prior to time-tagged detection by DET1 and DET2.
A histogram of the coincidences as a function of time difference between detector clicks was acquired with 300~ps binwidth and fitted to the second order correlation of a two-level system~\cite{redjem_single_2020}: 
\begin{equation}
    g^{(2)}(t) = b \left(1-(1-a)e^{-\frac{t}{\tau}}\right).
\end{equation}
The resulting second-order correlation of g$^{(2)}(0) = 0.38 \pm 0.08$ indicates single-photon emission. 
Poissonian error bars \edit{were} applied to the data points in Fig.~\ref{fig3}b. 
The error bar on g$^{(2)}(0)$ refers to the error of the fit calculated from the covariance in Python's \texttt{scipy.curve\_fit()} function. 

Emitter decay lifetimes were measured using pulsed 532~nm excitation at a \edit{repetition} rate of 34~MHz. 
A histogram of the SNSPD clicks as a function of \edit{the} delay time between the detection event and the pulsed laser trigger were first clipped to remove laser leakage and background (see SI Fig.~\ref{fig:lifetimefitting}a-b for details). 
The resulting decay curves fit well to a mono-exponential function:
\begin{equation}
    f(t) = a e^{-\frac{t-b}{\tau}}
\end{equation}
with an emitter lifetime of $\tau=8.21\pm0.14$~ns.
Here, the error bar on the lifetime of a single emitter is again calculated from the error of the fit. 

\subsection{System efficiency}
The collection efficiency of our system was characterized as follows: 
\begin{itemize}
    \item $\mathbf{\eta_\text{ec}}$: Simulated edge coupling efficiency between the cleaved waveguide mode and a perfectly aligned lensed fiber was simulated to be $8.3\%$ at 1280~nm.
    \item $\mathbf{\eta_\text{dipole-wg}}$: Maximum dipole coupling to a uni-directional waveguide mode was simulated to be $40\%$ (see SI Fig.~\ref{fig:effsims}a-b for details).
    \item $\mathbf{\eta_\text{filt}}$: Free space fiber filtering with two fiber collimators was measured to have $51.3\%$ transmission at 1280~nm.
    \item $\mathbf{\eta_\text{fiber}}$: Fiber routing to the two SNSPD cryostat detectors DET1 and DET2 was measured to be $90.6\%$ and $94.8\%$, respectively.
    \item $\mathbf{\eta_\text{splitter}}$: Insertion loss of fiber beamsplitter was measured to be $92\%$.
    \item $\mathbf{\eta_\text{det}}$: SNSPD detector efficiencies at 1280~nm for DET1 and DET2 were characterized to be $24\%$ and $21\%$, respectively. These SNSPDs were manufactured to have high efficiency at a target wavelength of 1550~nm.
\end{itemize}

The total system efficiency, given by 
\begin{equation}
    \eta_{\text{tot}} = \eta_{\text{ec}} \eta_{\text{filt}} \eta_{\text{fiber}} \eta_{\text{splitter}} \eta_{\text{det}}
\end{equation}
can, therefore, be upper bounded to $0.34\%$ for DET1 and $0.31\%$ for DET2. \edit{Using the above estimates for the collection efficiency and an observed saturated count rate of 1382$\pm$59~ counts/second under 78~MHz pulsed excitation, we estimate a lower bound on the quantum efficiency of $1\%$. However, the uncertainty on this estimate is large, due to low measured count rates and estimates of $\eta_\text{ec}$ and $\eta_\text{dipole-wg}$ being derived from simulation.}
\vspace{-10pt}

\section*{Acknowledgements}
The authors would like to acknowledge Dr. Matt Trusheim and Isaac Harris for useful discussions regarding emission physics and defect structure. The authors would also like to thank Dr. Dalia Ornelas-Huerta for contributions to the experimental setup. 

M.P. acknowledges funding from the National Science Foundation (NSF) Convergence Accelerator Program under grant No.OIA-2040695 and Harvard MURI under grant No.W911NF-15-1-0548.
C.E-H. and L.D. acknowledge funding from the European Union’s Horizon 2020 research and innovation program under the Marie Sklodowska-Curie grant agreements No.896401 and 840393.
I.C. acknowledges funding from the National Defense Science and Engineering Graduate (NDSEG) Fellowship Program and NSF award DMR-1747426.
D.E. acknowledges support from the NSF RAISE TAQS program.
This material is based on research sponsored by the Air Force Research Laboratory (AFRL), under agreement number FA8750-20-2-1007. 
The U.S. Government is authorized to reproduce and distribute reprints for Governmental purposes notwithstanding any copyright notation thereon.
The views and conclusions contained herein are those of the authors and should not be interpreted as necessarily representing the official policies or endorsements, either expressed or implied, of the Air Force Research Laboratory (AFRL), or the U.S. Government.

\edit{
\section*{Data availability}
The datasets generated and analysed during this study are available from the corresponding authors on reasonable request.
}

\bibliography{Si_artificial_atoms.bib}

\newpage
\onecolumngrid
\appendix
\renewcommand{\thefigure}{S\arabic{figure}}
\setcounter{figure}{0}
\newpage 

\section*{Supplementary Information}
\subsection{Measurement setup}
\label{sec:setup}

\begin{figure*}[hbtp]
  \centering
  \includegraphics[width=\textwidth]{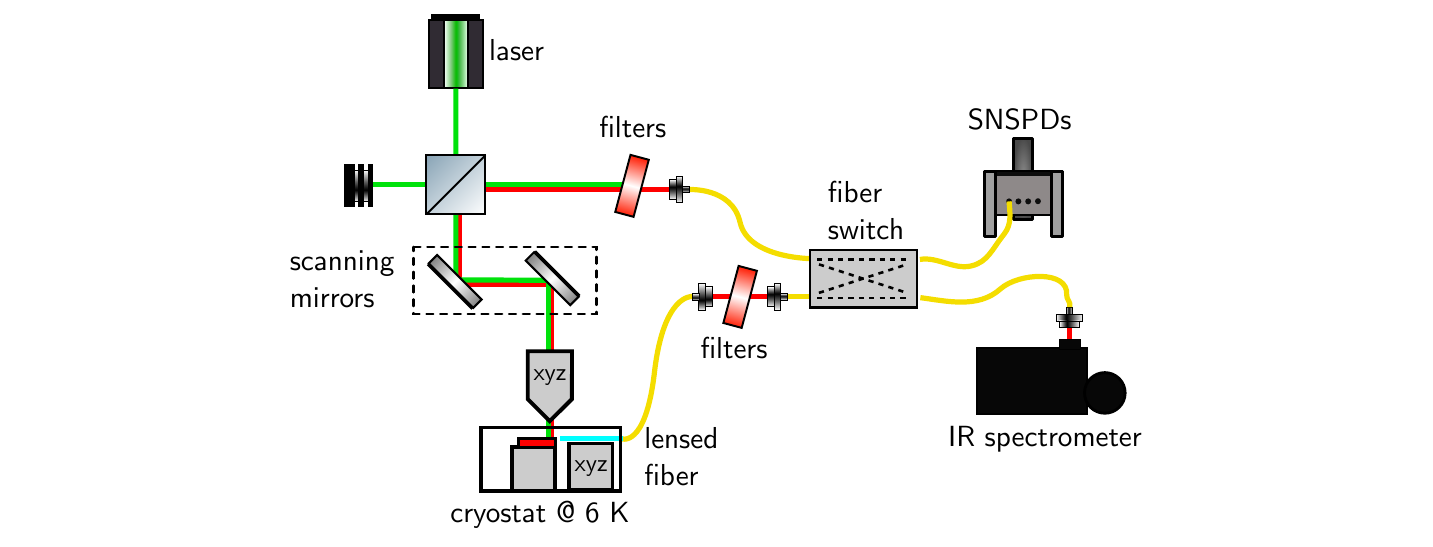}
\caption{\textbf{Measurement setup schematic.} Our setup combines confocal microscopy with fiber coupling. Laser light shines through a beam splitter and a set of galvanometers into an objective. For fiber-coupling, the PL from the waveguide couples into a aligned lensed fiber via free-space filters into a fiber switch leading to SNSPDs or an IR spectrometer. For confocal, PL from the sample reflects off the beam splitter via a filtering stage into the fiber switch.}
\label{fig:m1setup}
\end{figure*}

Figure~\ref{fig:m1setup} shows the measurement setup we used in this work.
Description of \edit{the} components of the measurement apparatus pertaining to the optical characterization of emitters is presented in the Methods section of the main text.
We include additional setup details in this section for completeness.
Our continuous wave excitation source is a Coherent Verdi G5 at 532~nm, and our pulsed excitation is \edit{an} NKT Photonics SuperK laser with a maximum repetition rate of 78~MHz \edit{and a pulse length below 1~ns}. 
To avoid background from our laser, the excitation path was filtered with a $532$~nm band-pass filter (Semrock Maxline bandpass with 2~nm FWHM).
Our microscope objective is a Mitutoyo $50\times$ M Plan APO and Mitutoyo $100\times$ M Plan APO SL.

Our cryostat is a Montana Instruments CR-057 with a home-made fiber feedthrough. 
The fiber we use (OZ Optics, spot size of 2.5~\textmu m at 1550~nm, and expected spot size of 2.1~\textmu m at 1300~nm) is mounted on a XYZ cryogenic piezoelectric stage (Attocube).
Our free-space filtering setup consists of a 1250~nm longpass (Thorlabs FEL1250) and a 1550~nm shortpass (Edmund Optics OD2) filters.
To detect our PL we use two superconducting nanowire single photon detectors (NIST) with detection efficiencies $21\%$ and $24\%$  \edit{and timing jitters below $172$ and $165$~ps}, readout with a Swabian Instruments Timetagger~20. 
Alternatively, we use a IR spectrometer consisting of a PyLon IR \edit{InGaAs linear} CCD \edit{array} from Princeton Instruments and \edit{two gratings with densities of 300~g/mm and 900~g/mm, 1.2~\textmu m and 1.3~\textmu m blazes, and resulting resolutions of 155~pm and 40~pm respectively.}
For the second order autocorrelation measurement we used a fiber beam splitter (Thorlabs TW1300R5F1) after the filtering station.

\subsection{Coupling efficiency simulations between emitter, waveguide, and fiber}
\label{sec:coupling}

\begin{figure*}[hbtp]
  \centering
  \includegraphics[width=\textwidth]{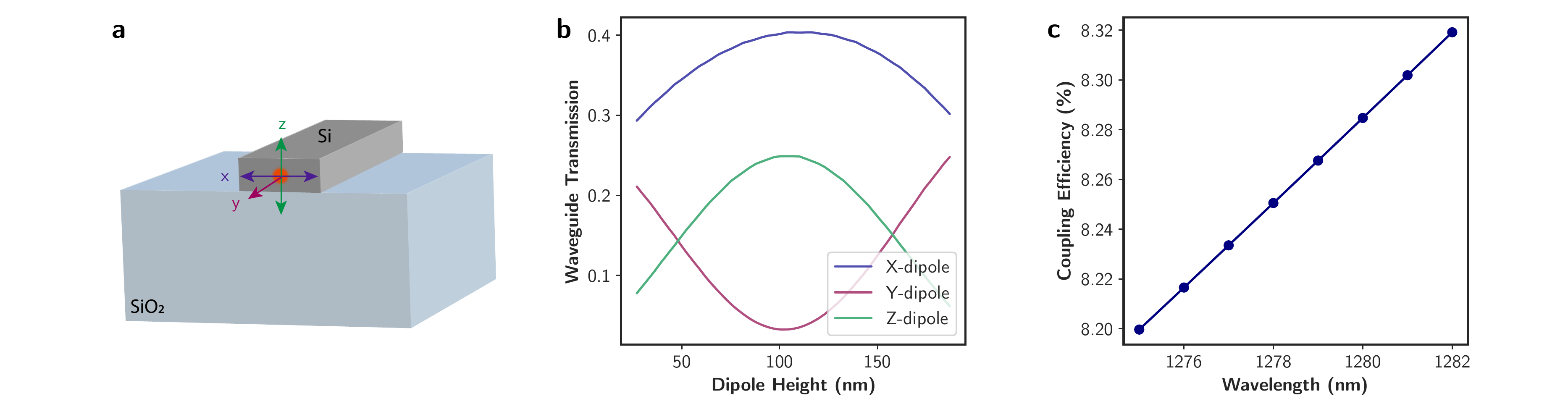}
\caption{\textbf{Coupling efficiency simulations.} a) Schematic showing the geometry and the dipole orientations simulated for b) the uni-directional waveguide coupling results. c) Coupling efficiency simulated via modal overlap between our lensed fiber mode and our cleaved waveguide facet.}
\label{fig:effsims}
\end{figure*}

Finite difference time domain (Lumerical FDTD) simulations were performed to calculate the emitter-waveguide coupling as a function of \edit{the} emitter height in the waveguide. 
Directional waveguide-coupled dipole emission was measured using a power monitor at one end of the waveguide, with a dipole emitter oriented in either the X-, Y-, or Z-direction positioned at a range of heights within the waveguide (Fig.~\ref{fig:effsims}a).
The single-ended waveguide transmission for each direction and dipole height is shown in Fig.~\ref{fig:effsims}b.
We observe \edit{an} improved waveguide coupling for dipoles oriented transverse to the waveguide axis and near the center of the silicon film.
Our asymmetric waveguide geometry preferentially supports \edit{quasi-}TE modes over \edit{quasi-}TM, so the improved coupling for \edit{the} dipoles oriented along the long transverse axis of the waveguide agrees with expectations. 
We note that \edit{the} waveguide collection could be improved by up to a factor of two by adding a reflector at the other end of the waveguide. 

\edit{The} edge coupling efficiency between the cleaved waveguide mode (shown in Fig~\ref{fig1}f) and a perfectly aligned lensed fiber was simulated using Lumerical MODE Solutions eigenmode solver.
We simulated the mode at the focus spot of the fiber as a Gaussian beam with 2.1~µm $1/e^2$ diameter, and the waveguide mode as the output of a waveguide mode eigenmode simulation.
\edit{Our} mode overlap calculations\edit{, including the Fresnel reflection at the waveguide-air interface,} yield the coupling efficiency in Fig.~\ref{fig:effsims}c, and a $8.25\%$ efficiency at 1280~nm.

\subsection{Confocal collection from vertically-coupled structures}
\label{sec:confocal}

\begin{figure*}[!hbtp]
  \centering
  \includegraphics[width=\textwidth]{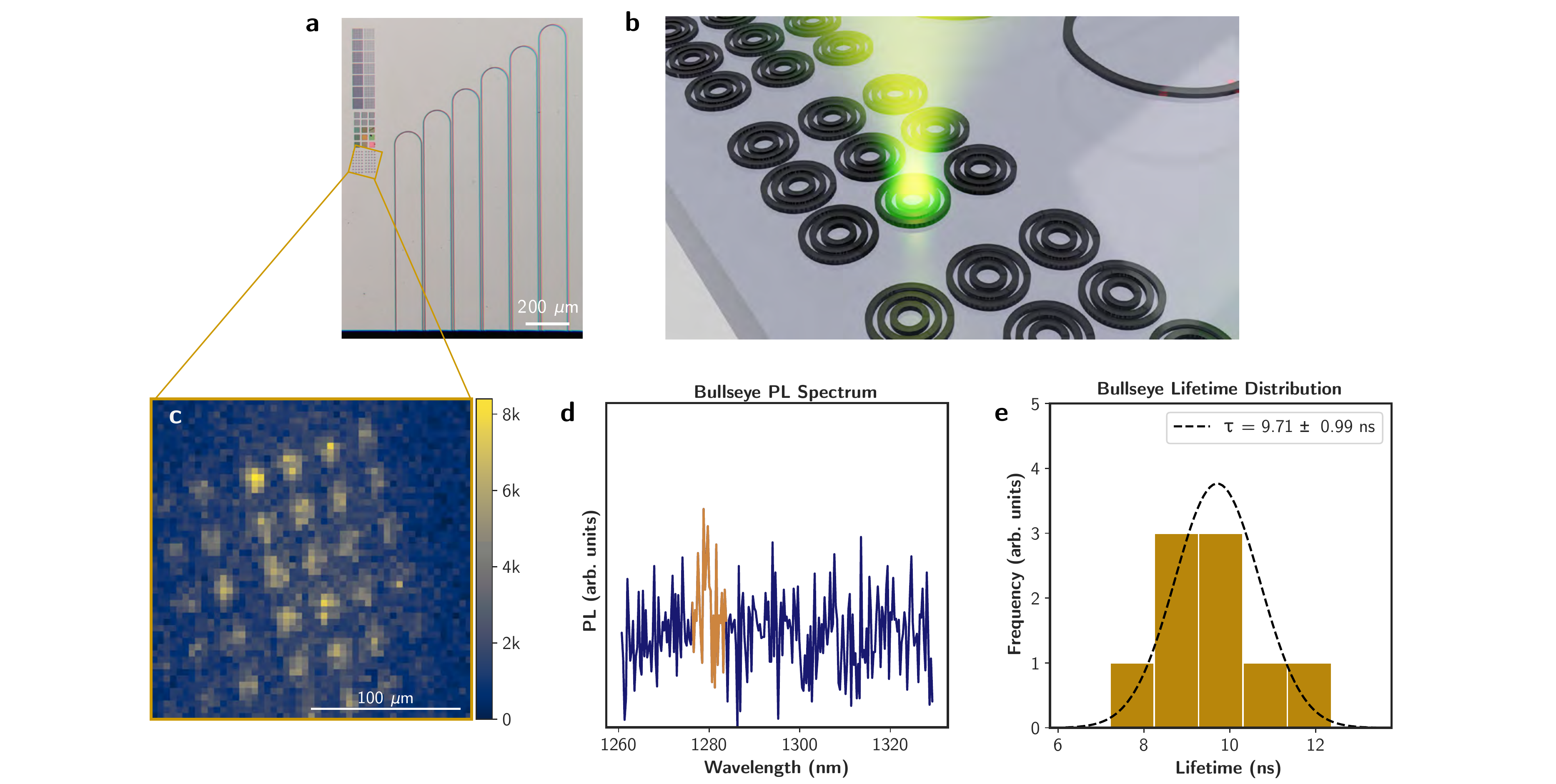}
\caption{\textbf{Confocal collection of G-centers from circular grating bullseye structures \edit{at 6~K}.} a) \edit{A} sample micrograph and b) \edit{a} schematic of the bullseye structures used to enhance vertical coupling of G-centers. c) \edit{The} PL map shows emission from the structures. d) PL signal from one of the devices. e) Lifetime statistics for 9 devices.}
\label{fig:confocal}
\end{figure*}

In addition to waveguides, the fabricated G-center sample consists of bulls-eye grating structures, designed for a vertically-coupled cavity mode near 1275~nm (Fig.~\ref{fig:confocal}a-b) with $Q \sim 500$. 
The objective used for both excitation and collection was a Mitutoyo $50\times$ M Plan APO long working distance objective (NA=0.55). 
It is worth noting that the specified objective has an designed operating range of 435-655~nm, which is optimized for transmission of the 532~nm excitation path. 
However, confocal collection around 1280~nm falls outside this range.
Raster scans of the vertically-coupled PL signal were acquired in the same way as the through-waveguide measurements. 
Distinct PL intensity maxima are observed over the bulls-eye structures (Fig.~\ref{fig:confocal}c); however, \edit{the} observed count rates were much lower for the same excitation power than with through-waveguide emission. 
A background-corrected PL spectrum from one of these structures was collected with the maximum spectrometer integration time of 800 seconds. 
A dim peak at 1278.622~nm \edit{was} observed (Fig.~\ref{fig:confocal}d), suggesting the presence of the G-centers in these structures. 
Lifetimes for 9 of these structures were measured, using the same protocol for fitting as the waveguide samples. 
The distribution of the lifetimes in the bulls-eyes was observed to be $9.71 \pm 0.99$~ns, slightly higher than \edit{those} observed for G-centers in waveguides. 

\subsection{Ion implantation simulations}
\label{sec:implant}
\begin{figure*}[htbp]

  \centering
  \includegraphics[width=\textwidth]{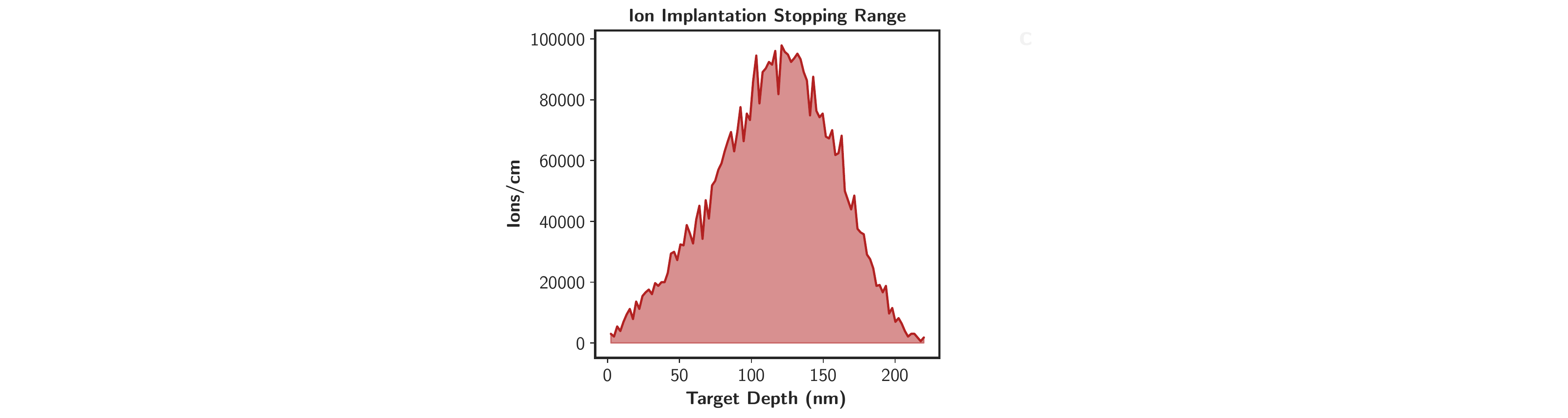}
\caption{\textbf{Ion implantation depth simulation.} \edit{A} simulated mean ion depth of $113.3$~nm and \edit{a} standard deviation of $41.3$~nm was observed.}
  \label{fig:srim}
\end{figure*}

\edit{The} carbon ion implantation depth was simulated using Stopping and Range of Ions in Matter (SRIM).
The simulated layer was 220~nm of $^{28}$Si with a density of $2.3212$~g/cm$^3$, and the simulated ions were $^{12}$C with an acceleration of 36~keV and incident at an angle of $7^\circ$.
The simulation results, shown in Fig.~\ref{fig:srim} yield an ion range (i.e. mean ion depth) of $113.3$~nm with a straggle (i.e. standard deviation) of $41.3$~nm, a skewness of -0.259 and a Kurtosis of 2.6411.

\subsection{Background correction for saturation}
\label{sec:background}

\begin{figure*}[hbtp!]
  \centering
  \includegraphics[width=\textwidth]{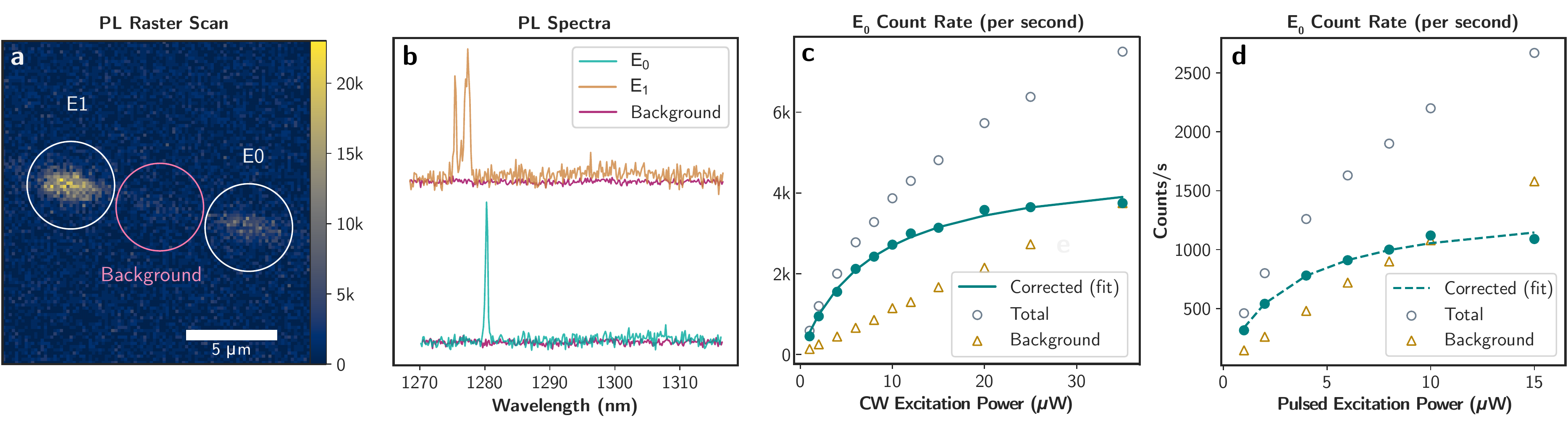}
\caption{\textbf{Background subtraction protocol for G-center through-waveguide measurements \edit{at 6~K}.} a) \edit{A} PL raster scan showing emitter and background regions of the waveguide. b) \edit{A} PL spectra for the three regions indicated in (a). c) Background-subtracted saturation curve for CW 532~nm excitatation. d) Background-subtracted saturation curve for pulsed 532~nm excitation.}
\label{fig:backgroundsubtraction}
\end{figure*}

A \edit{PL} signal was observed in the waveguide regions between PL maxima, particularly at high excitation powers. 
As a result, a protocol to measure and subtract this background was necessary to separate the count rate from the emitters from the background counts generated in the waveguide. 
First, PL maps were taken to isolate points that contain G-centers, as well as \edit{to} locate a point in the waveguide to measure the background. 
The regions isolated for this measurement are shown in Fig.~\ref{fig:backgroundsubtraction}a. 
\edit{The} spectra of the PL signal from the two emitter points, E$_{0}$ and E$_{1}$ show ZPL peaks near 1280~nm, whereas the background waveguide region shows no ZPL emission (Fig.~\ref{fig:backgroundsubtraction}b). 
The count rate at the single emitter spot E$_{0}$ and the background region was measured over a range of excitation powers.
The background-corrected emitter saturation was calculated by subtracting the total count rate at the emitter from the counts measured in the waveguide background region. 
Figures~\ref{fig:backgroundsubtraction}c-d show the results of the background correction for both CW and pulsed 532~nm excitation. 

\subsection{Lifetime fitting}
\label{sec:lifetime}

\begin{figure*}[hbtp!]
  \centering
  \includegraphics[width=\textwidth]{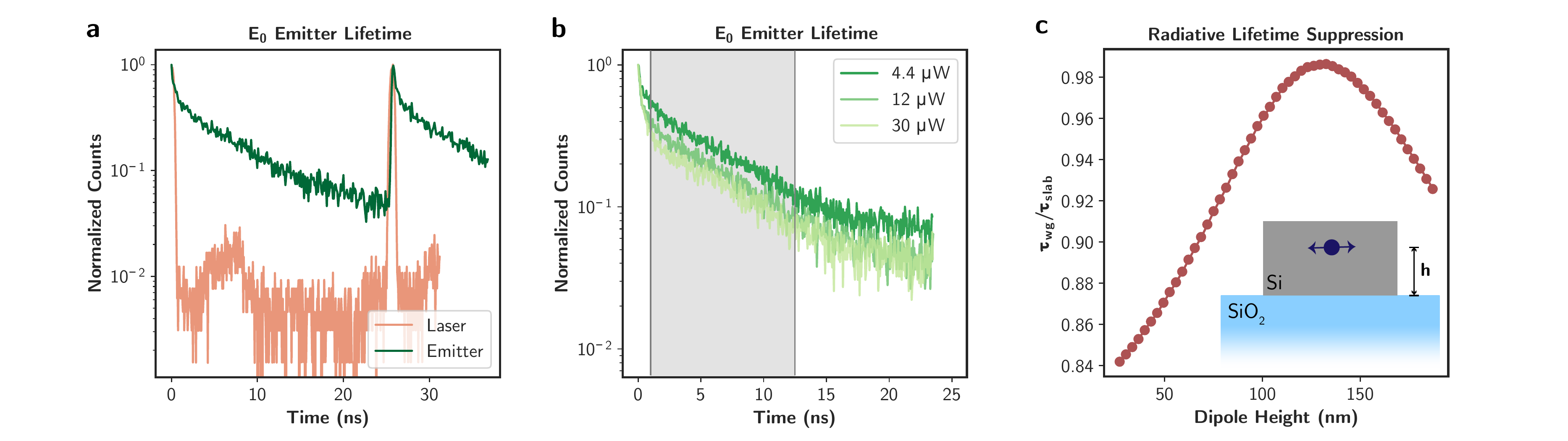}
\caption{\textbf{Emitter lifetime fitting and simulations.} a) Lifetime measurement comparative between an emitter and the green laser, highlighting the need to time-clip the signal. b) \edit{The} emitter lifetime measurements under varying excitation power yield negligible variation for an order of magnitude power variation. c) \edit{Our} simulations of Purcell enhancement in a waveguide versus in a SOI slab yield a nearly-negligible effect on the expected lifetime ratio. \edit{The measurements in a) and b) were performed at 6~K.}}
\label{fig:lifetimefitting}
\end{figure*}

\edit{The} emitter lifetimes were measured using 532nm pulsed excitation from the SuperK laser. 
A repetition rate of 34~MHz was selected to enable decay curves of up to 29~ns to be acquired.
The resulting raw decay curves exhibit both an initial peak at the onset of the pulse, as well as flattening towards the end of the pulse period from count background. 
Fig.~\ref{fig:lifetimefitting}a shows a plot of the normalized emitter lifetime compared to the normalized lifetime of the pulsed laser. 
A bi-exponential fit of the raw lifetime data confirms that the time constant of the initial peak matches with that of the excitation laser. 
Additionally, it can be seen in Fig.~\ref{fig:lifetimefitting}b that increasing the excitation power does not change the lifetime slope; however, the fractional contribution of the initial peak to the total counts increases with higher laser power. 
Furthermore, it is observed that the decay becomes background-limited near 15~ns post-excitation. 
The final lifetime is calculated by clipping the raw lifetime to a range between 1~ns and 12.5~ns before fitting to a single exponential function (Fig.~\ref{fig:lifetimefitting}b).

\edit{We compared our} simulations of the local density of optical states (LDOS) for a dipole emitter in our waveguide (400~nm $\times$ 220~nm) to \edit{those of} the dipole LDOS in \edit{a} 220~nm thick slab SOI to estimate the change in the radiative lifetime of our patterned sample. 
\edit{The} dipole radiation power was simulated in Lumerical FDTD along the three cardinal axes to calculate the total LDOS for an emitter \edit{with a} random orientation within the waveguide. 
A similar simulation was calculated for dipoles placed in \edit{a} slab SOI. 
The radiative lifetime, $\tau$, is inversely proportional to the LDOS, $\rho$, so the suppression in radiative lifetime in the waveguide compared to a slab \edit{is}
\begin{equation}
    \frac{\tau_{\text{wg}}}{\tau_{\text{slab}}} = \frac{\rho_{\text{slab}}}{\rho_{\text{wg}}}
\end{equation}
\edit{and} was calculated over a range of possible dipole heights within the waveguide (Fig.~\ref{fig:lifetimefitting}c). 
From this simulation, we observe that the change in radiative lifetime due to waveguide patterning of a slab does not fully explain the difference observed between our results and that of isolated single G-centers in SOI slabs~\cite{redjem_single_2020}.

\subsection{Simulations of strain distribution in SOI waveguides and slabs}
\label{sec:inhomstrain}
It is well known that commercial SOI wafers have significant intrinsic stress.
The presence of a defect in the lattice induces \edit{stress relaxation} into strain gradients localized around the defect, which may result in a significant inhomogeneous ditribution for a constrained SOI slab, such as \edit{for} the results presented in~\cite{redjem_single_2020}. 
In contrast, for the more \edit{mechanically} compliant case of a patterned waveguide, the stress in the film relaxes largely into homogeneous strain (i.e. by deforming the waveguide cross-section), potentially resulting in a narrower inhomogeneous distribution.

\begin{figure*}[hbtp!]
  \centering
  \includegraphics[width=\textwidth]{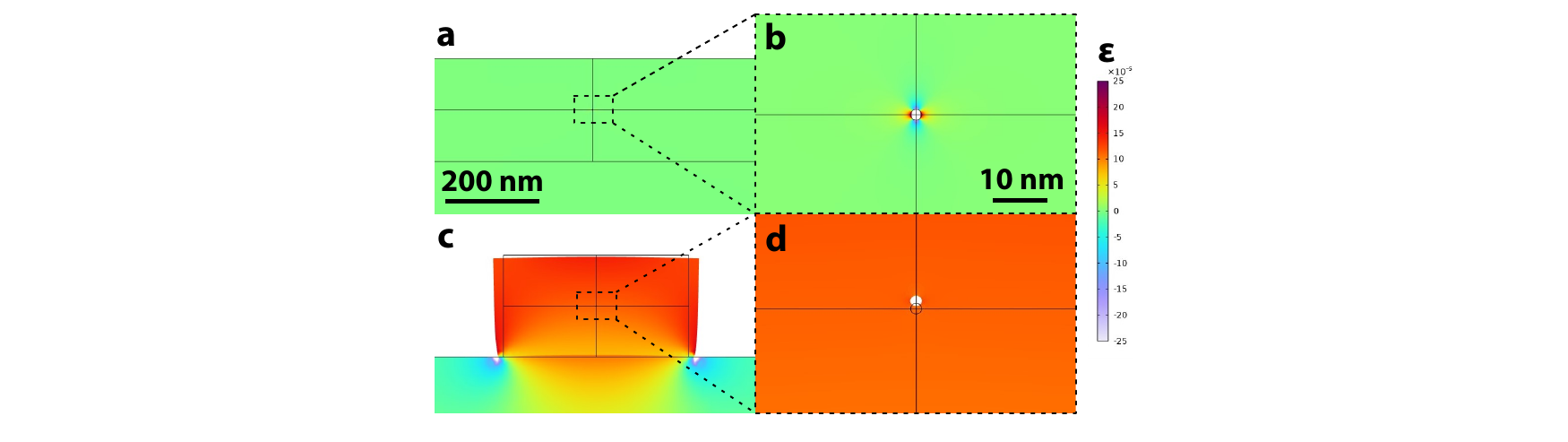}
\caption{\textbf{Simulations of strain distribution for a slab and a waveguide.} a) The strain distribution in a SOI slab. b) \edit{The} zoom-in shows large strain gradients concentrated around the defect. c) The waveguide-patterned structure relaxes its stress by strain. d) The zoom-in around the defect shows up to an order of magnitude reduced strain gradients.}
\label{fig:strainsim}
\end{figure*}

We qualitatively model\edit{ed} this effect by assuming a nanoscale air defect (simulated as an air gap of 1~nm radius) in a isotropic silicon film.
Using COMSOL Multiphysics v5.5, we simulated a 2D silicon slab with fully clamped sidewalls and our defect under the 38~MPa biaxial (x-axis in the simulation) compressive stress reported for photonic SOI.
The resulting strain distribution is shown in Fig.~\ref{fig:strainsim}a and b, and shows height-independent volumetric strain gradients in the order of $\varepsilon_\text{max}=20\times10^{-5}$ in the vicinity of the defect.
In contrast, the simulation results for a 400~nm wide waveguide in Fig.~\ref{fig:strainsim}c and d show reduced strain gradients for all defect positions ($\varepsilon_\text{max}=10\times10^{-5}$) and an order of magnitude ($\varepsilon_\text{max}=1.5\times10^{-5}$) in the center of the waveguide.

Although these results do not quantitatively model a real defect in a solid, they provide a qualitative intuition for a possible explanation to our observation of a reduced inhomogeneous distribution in waveguides.

\subsection{Defect-related carrier dynamics}
\label{sec:rates}
\begin{figure*}[hbtp!]
  \centering
  \includegraphics[width=\textwidth]{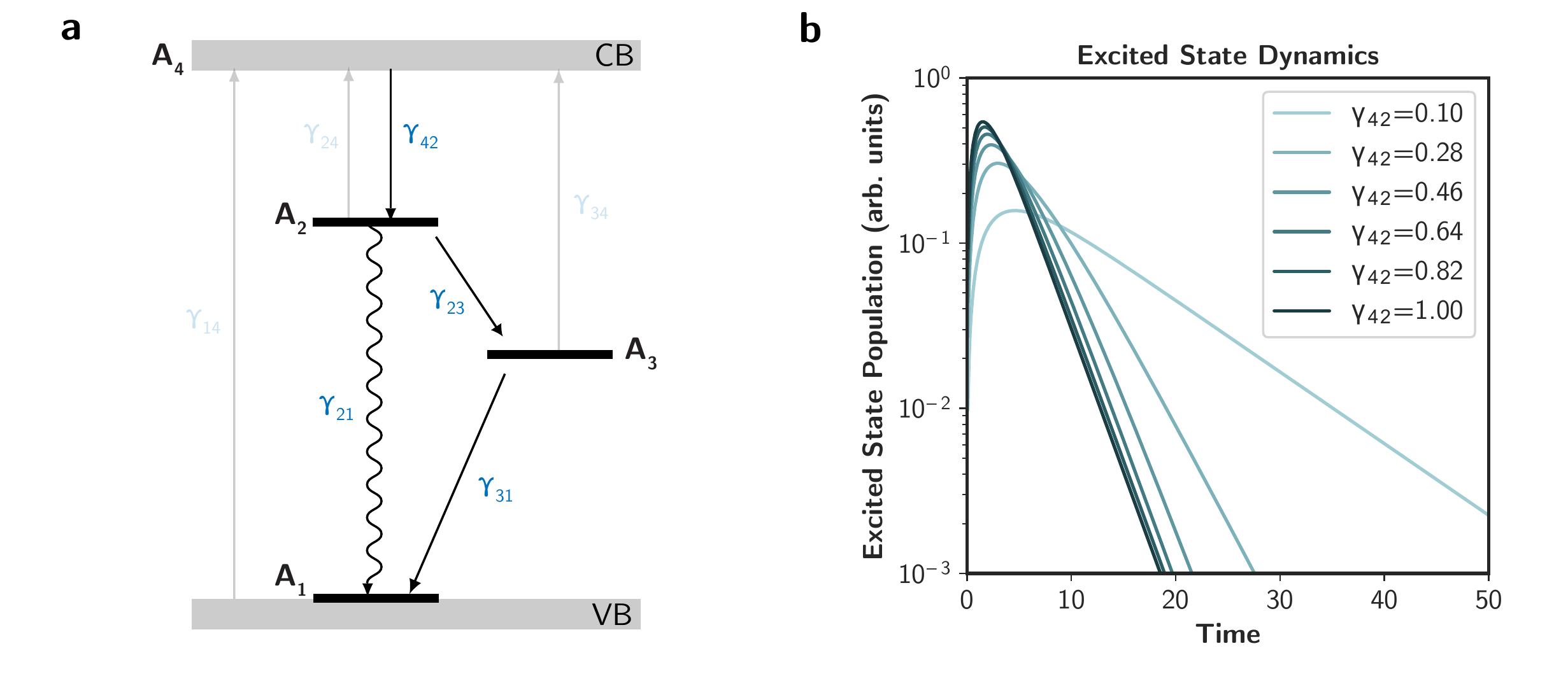}
\caption{\edit{\textbf{Energy level description and rates.}} a) Energy levels and transitions for defect-assisted recombination in G-centers. A$_4$=CB=conduction band, VB=valence band, A$_2$-A$_1$ is the G-center ZPL transition, and A$_3$ is the metastable state. Note that here we made A$_1$ degenerate with VB as described in Ref.~\cite{udvarhelyi_identification_2021}. b) Excited state dynamics show a decrease in lifetime under increased electron capture rates $\gamma_{42}$.}
\label{fig:rate_equations}
\end{figure*}

The photodynamics of the G-center emission under above-bandgap excitation is described here with the rate equation model depicted in Fig.~\ref{fig:rate_equations}.
As silicon possesses an indirect bandgap at a transition near $1100$~nm (1.17~eV), we assume that the band-to-band radiative recombination rate is negligible. 
The primary mechanisms governing carrier dynamics in silicon are Shockley-Read-Hall (SRH) trap recombination, surface recombination, and Auger recombination~\cite{coldren_diode_2012}.
The waveguide devices presented in this paper do not have a top oxide cladding and are not passivated to reduce surface trap states; therefore, we expect a non-negligible recombination from surface effects at the waveguide sidewalls. 
The main pathway to populate the excited state of the G-center is determined by the rate $\gamma_{42}$, which is an effective rate including the influence of trap states in the silicon.
Once populated, the emission dynamics from the G-centers is following the behaviour of a 2 level system with an additional shelving state.


The dynamics of this system are described by the following rate equations, where $N_{i}$ corresponds to the carrier occupation density at level $A_{i}$.

\begin{eqnarray}
    \frac{d}{dt}
    \begin{bmatrix}
    N_1 \\
    N_2 \\
    N_3 \\
    N_4
    \end{bmatrix}
    = 
    \begin{bmatrix}
    -\gamma_{14} & \gamma_{21} & \gamma_{31} & 0 \\
    0 & -(\gamma_{24}+\gamma_{23}+\gamma_{21}) & \gamma_{32} & \gamma_{42} \\
    0 & \gamma_{23} & -(\gamma_{31}+\gamma_{32}+\gamma_{34}) & 0 \\
    \gamma_{14} & \gamma_{24} & \gamma_{34} & -\gamma_{42}
    \end{bmatrix}
    \begin{bmatrix}
    N_1 \\
    N_2 \\
    N_3 \\
    N_4
    \end{bmatrix}
    \label{eqn:rateequations}
\end{eqnarray}

The transition rates $\gamma_{ij}$ from state A$_i$ to state A$_j$ depend on a number of material properties in the patterned silicon structures, and as such, \edit{the} calculation of \edit{the} exact rates is out the scope of this study. 
The level occupation densities can be obtained using a matrix exponential solution to the system of coupled differential equations. 
Due to the coupled nature of the rate equations, the time-dependent state occupations depend on contributions from each of the rate terms, determined by the eigenvalues of the coefficient matrix in Eqn.~(\ref{eqn:rateequations}).
\edit{The} assumptions and general physical mechanisms that can govern these rates are described below: 
\begin{itemize}
    \item $\gamma_{14}$: denotes the electron pumping rate, under the assumption that the conduction band occupation is much lower than the valence band occupation ($N_1 >> N_4$). Our lifetime measurements are performed under pulsed excitation, so we assume pumping and re-pumping processes (including $\gamma_{24}$ and $\gamma_{34}$) to the conduction band to be zero after the initial pulse. 
    \item $\gamma_{24}$: the rate of thermalized excitation of electrons from the upper defect level to the conduction band (a function of material doping and temperature, $T$), which includes a re-pumping term (set to zero after the pulse, see above).
    \item $\gamma_{42} \propto \sigma_e N_{\text{surface}} N_{\text{carrier}}$: the rate of electron capture from the conduction band to the upper trap state depends on material properties of the wafer and structures, including the electron capture cross-section, $\sigma_e$; the density of surface traps, $N_{\text{surf}}$; and the carrier density, $N_{\text{carrier}}$.
    \item $\gamma_{21} \propto \frac{F_{\text{Purcell}}}{\tau_{\text{rad}}}$: describes effective radiative emission rate from the defect excited state to ground state in the presence of a perturbed local density of optical states (LDOS), where enhancement in the spontaneous emission rate is quantified by the Purcell factor
\end{itemize}

After laser excitation, we assume all the carrier population in N$_4$ (i.e. the conduction band) and let the system of equations evolve over time.
Figure~\ref{fig:rate_equations}b shows the evolution of the carrier density in the excited state A$_2$ in this model while varying the rate of electron capture to our defect, $\gamma_{42}$.
We observe that an increased $\gamma_{42}$ leads to shorter lifetimes.

\edit{The measured G-center lifetimes reported in the literature vary between 4.5~ns~\cite{baron_single_2022, beaufils_optical_2018} and 35.8~ns~\cite{redjem_single_2020}.
There are many possible physical mechanisms that can result in variations in the carrier dynamics, including band bending from surface charges or from compressive strain in the silicon wafer, wafer-to-wafer variations in silicon thickness and doping, or opportunistic emission from other defect centers. 
Correcting for these sample-dependent variations, perhaps by using techniques such as surface passivation, will be a critical step towards building a scalable artificial atom platform in SOI. 
However, investigation of these mechanisms is out of the scope of the current study. 
}

\subsection{PL stability measurements}
\label{sec:stab}

\begin{figure*}[hbtp!]
  \centering
  \includegraphics[width=\textwidth]{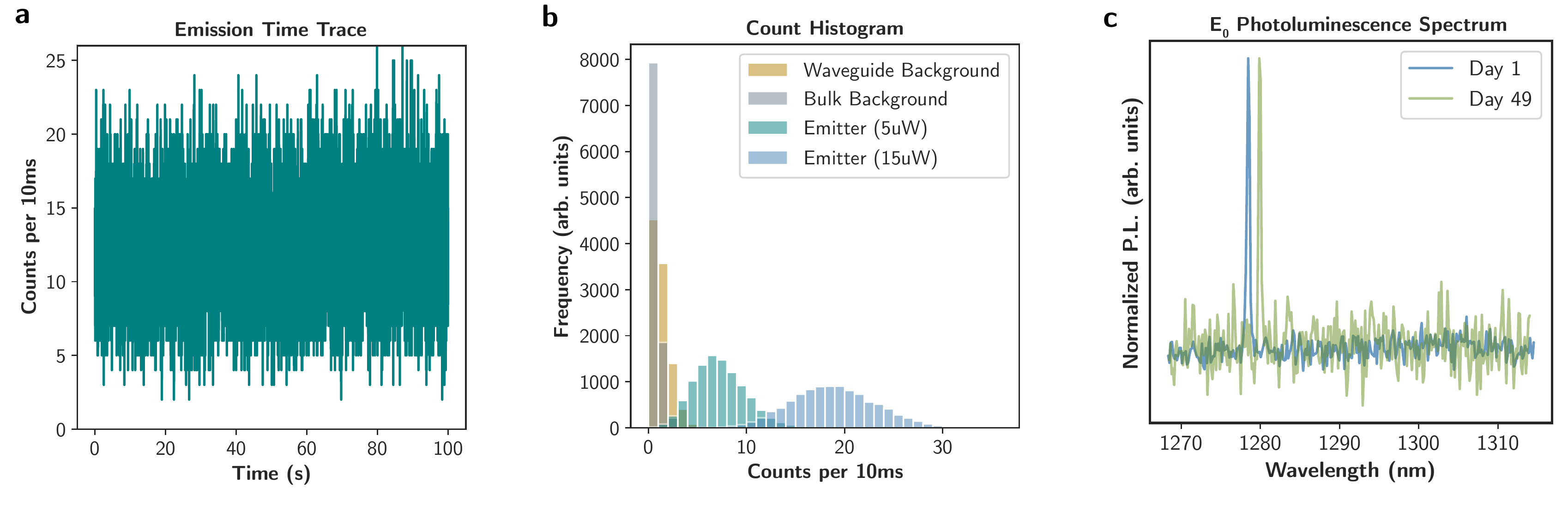}
\caption{\textbf{PL emission stability measurements \edit{at 6~K}}. a) Time trace showing stable PL emission from a single G-center for 100 seconds at 10~\textmu W 532nm continuous-wave excitation. b) \edit{The} comparative histograms show that the frequency of counts for a 10~ms time bin width at 5~\textmu W excitation fits a Poisson distribution. \edit{c) Normalized photoluminescence spectra collected from a single emitter with acquisition separated by 49 days. Note that the apparent ZPL wavelength shift is largely caused by the re-calibration of our spectrometer in between those measurements.}}
\label{fig:stab}
\end{figure*}

We investigated the stability of our waveguide-coupled G-centers via time-tagged PL collection.
Figure~\ref{fig:stab}a shows a time trace of PL emission from a single G-center during a 100~s acquisition period with a binning time of 10~ms.
Figure~\ref{fig:stab}b shows comparative histograms for \edit{the} emitter and \edit{the} waveguide background under a range of excitation powers.
Our emitter histograms fit to Poisson distributions, suggesting non-blinking and stable behavior at timescales longer than 10~ms.
Finally, Figure~\ref{fig:stab}c shows that spectral resolution-limited PL can be observed from a single emitter in measurements separated by over a month and following multiple cryostat cooldown cycles. 
The spectral shift in the ZPL peak between the two measurements is likely due to recalibration of the spectrometer.

Our stability measurements are limited to a 10~ms time bin due to our limited counts arising from our low system efficiency. 
Higher coupling efficiencies are required to probe emitter stability at faster timescales.

\subsection{\edit{Optical trimming of silicon color centers}}
\label{sec:tuning}

\edit{
Our trimming experiments were performed in-situ in our cryogenic setup described in SI Section~\ref{sec:setup} without additional modifications.
Our trimming protocol consisted of:
\begin{enumerate}
    \item Localization of emitters by PL maps using 25~\textmu W (estimated power density of $13.6$~kW$\cdot\text{cm}^{-2}$) of 532~nm CW laser excitation and our SNSPD detectors, followed by spectrometry (900~gr/mm grating, 40~pm resolution) under the same excitation conditions.
    \item Local irradiation with a 532~nm CW laser with powers in the order of 100~\textmu W (estimated power density of $54.4$~kW$\cdot\text{cm}^{-2}$) for 15~s.
    \item Characterization of the emitter by PL maps and spectrometry as in the first step.
    \item Repetition of steps 2-3 with increasing irradiation power.
\end{enumerate}
}

\begin{figure*}[hbtp!]
  \centering
  \includegraphics[width=\textwidth]{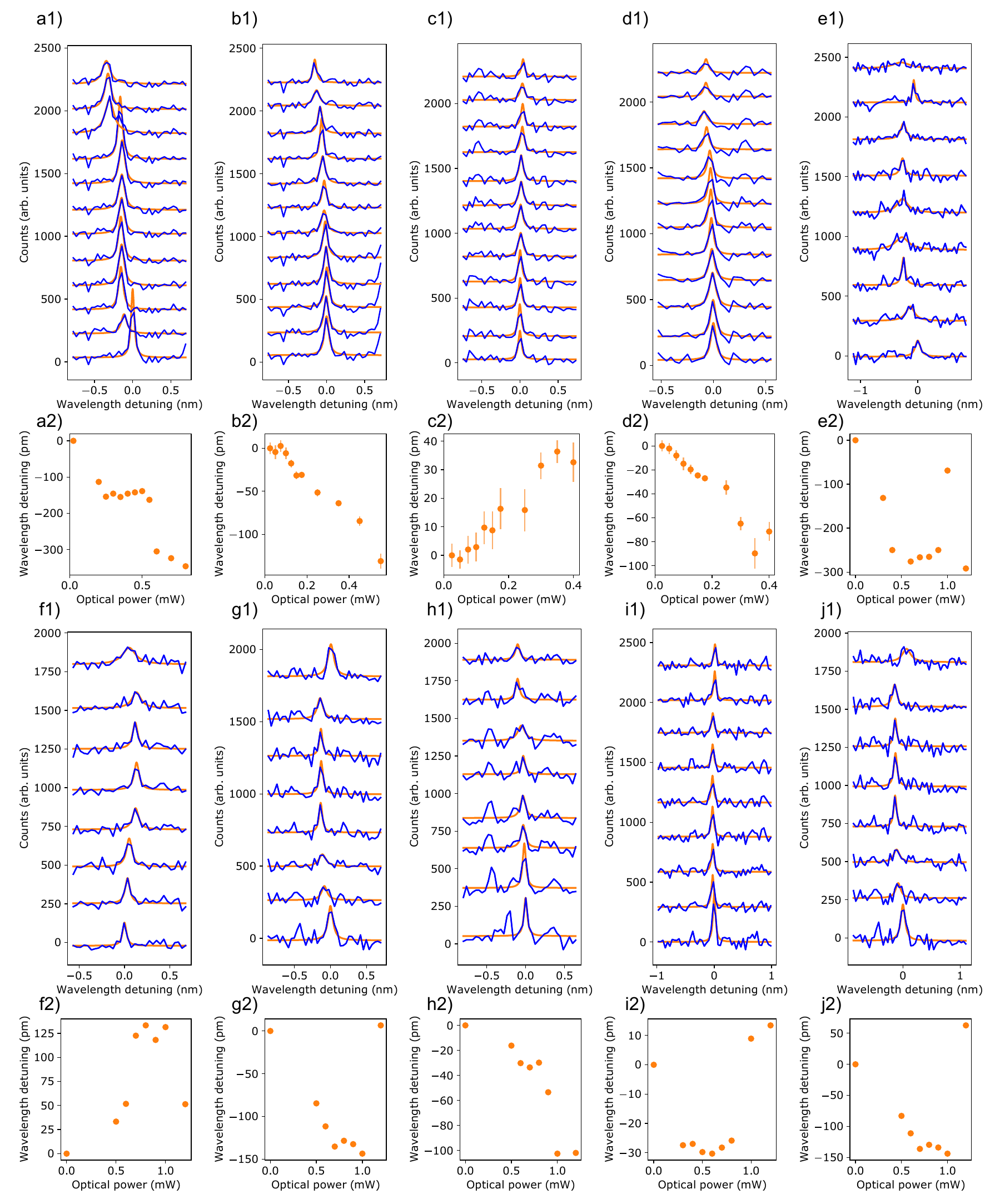}
\caption{\edit{\textbf{Trimming results for all the tested devices at 6~K.} Each vertically stacked pair of subfigures (e.g. a1) and a2)) are from the same emitter. 1) Spectra under increasing irradiation power (blue, increasing power from bottom to top within each subfigure), and Lorentzian fitting (orange). 2) Fitted central wavelength shift versus irradiated optical power.}}
\label{fig:tuningall}
\end{figure*}

\begin{figure*}[hbtp!]
  \centering
  \includegraphics[width=\textwidth]{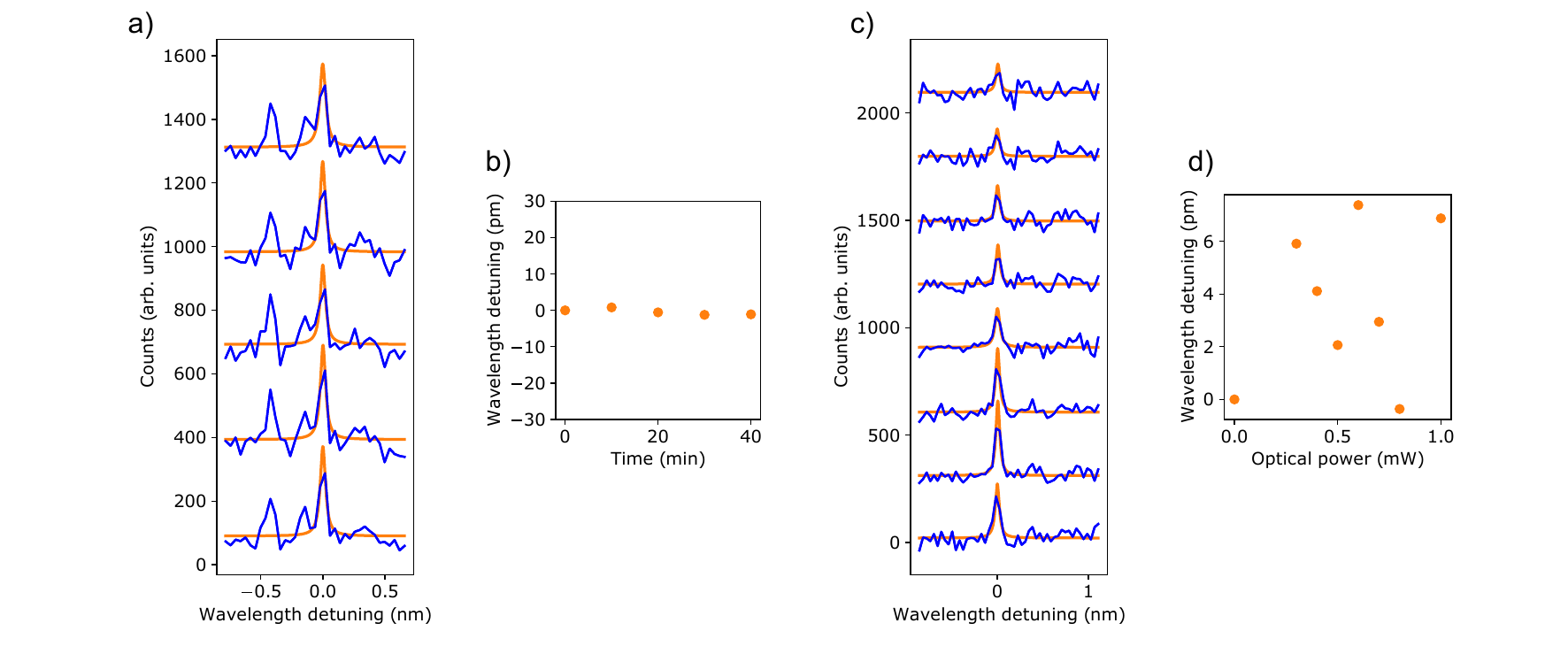}
\caption{\edit{\textbf{Stability of trimmed emitters at 6~K.} a) The stability of the PL spectra of a small ensemble of emitters (where b) shows a fit to one of the emitters) monitored after trimming demonstrates the non-volatility of our method. c) Fitted spectra and d) central wavelength for the only non-trimmable emitter within a sample of 12.}}
\label{fig:tuningstab}
\end{figure*}

\edit{
We performed our trimming characterization protocol in 12 bright emitters (5000 to 10000~cts/s on our SNSPDs with a 1300-1550~nm bandpass filter) under PL excitation.
From those, we observe trimming in 11 emitters, with spectral measurements and Lorentzian fits shown in Fig.~\ref{fig:tuningall} and Fig.~\ref{fig:tuningstab}b and c.
We observe spectral blue shifts in 9 out of the 11 shifting emitters, while two of them red shift, including emitters in the same spatial location shifting in opposite directions. 
The maximum magnitude of the trimming is 300~pm (55~GHz) and depends on the emitter, with a relatively constant bandwidth limited by our spectrometer resolution.
When nearing 1~mW of irradiation (estimated power density of $544.3$~kW$\cdot\text{cm}^{-2}$), some of the emitters appear to shift back to near their original spectral position.
Consecutive measurements without irradiation at any step in the process did not show spectral shift (see Fig.~\ref{fig:tuningstab}a and b for continuous monitoring of a small ensemble over 40~min).
The emitter that did not show shifts showed instead diminishing brightness with increasing irradiation power, and is plotted in Fig.~\ref{fig:tuningstab}c and d. 
We note that we did not observe an increase in the cryostat temperature during our measurements.
}

\begin{figure*}[hbtp!]
  \centering
  \includegraphics[width=\textwidth]{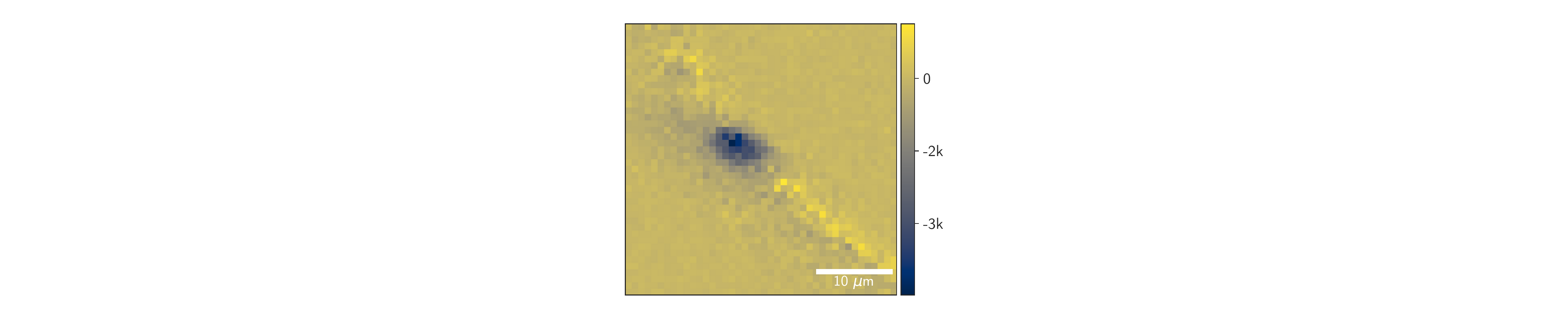}
\caption{\edit{\textbf{Deactivation of emitters with high optical powers.} The subtracted PL map from Figs.~\ref{fig:tuning}d and e show localized deactivation of single emitters and small clusters.}}
\label{fig:deactivation}
\end{figure*}

\edit{
For higher powers, we observe local deactivation of emitters.
Figure~\ref{fig:tuning}d shows a PL map of a section of our waveguide containing several emitters, with a brighter ensemble in the center where the irradiation was focused.
For this location, after 0.9~mW (estimated power density of $489.9$~kW$\cdot\text{cm}^{-2}$)  of local irradiation for 60~s, the small ensemble in the center is deactivated, as shown in Fig.~\ref{fig:tuning}e, without deactivating nearby emitters. 
This is evidenced by the subtracted PL map in Fig.~\ref{fig:deactivation}, showing the location where the emitter was deactivated.
The faint mark following the waveguide in Fig.~\ref{fig:deactivation} is a result of a slight misalignment between the first and second PL maps.
}

\begin{figure*}[hbtp!]
  \centering
  \includegraphics[width=.4\textwidth]{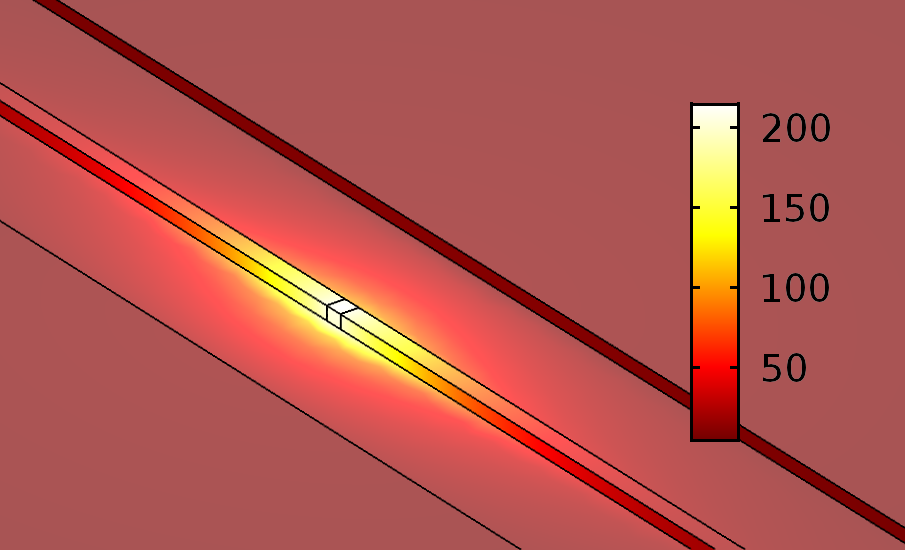}
  \includegraphics[width=.4\textwidth]{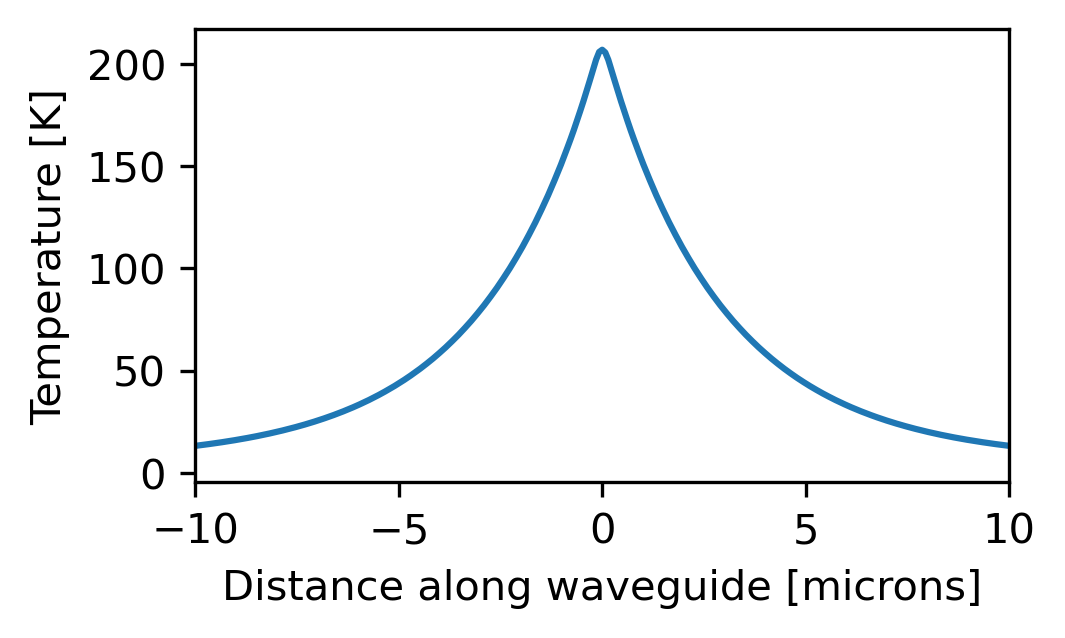}
\caption{\edit{\textbf{Simulation of local temperature increase in an irradiated waveguide.} 
A 3D plot of the simulated system (left), including the Si waveguide and SiO$_2$ bottom cladding. The color bar represents temperature in K. The white cube in the center acts as the heat source in our simulation. 
A 1D slice along the center of the waveguide (right) shows a full-width-half-maximum of $\sim$5~\textmu m.}}
\label{fig:thermcond}
\end{figure*}

\edit{
Local annealing of point defects in the waveguide could explain these observations. 
To explore this hypothesis, we simulate the thermal performance of our waveguide system subjected to localized heating. 
We assume 100~\textmu W of heat incident upon a 300~nm$\times$300 nm square at the top of the waveguide.
This corresponds to order 1~mW of 532~nm excitation: roughly a third of this power is reflected by the silicon-air interface; a tenth is assumed to be absorbed; and the remainder is absorbed in a less localized manner or reflected. 
The assumed absorption is estimated from silicon's room-temperature optical attenuation coefficient of $\sim$1/\textmu m at 532~nm~\cite{green_opticalsi_2008}; there is evidence that the absorption coefficient at telecommunication wavelengths is roughly constant down to low temperatures~\cite{degallaix_opticalsicryo_2014}.
We use thermal conductivity estimates at 5~K for thin silicon (10~W/m/K) and silicon dioxide (0.1~W/m/K) from Ref.~\cite{asheghi_cryothermsoi_1998} and the same for the bulk silicon (100~W/m/K) substrate from Ref.~\cite{glassbrenner_thermalsi_1964}. 
Note that these thermal conductivities increase with temperature for the temperature ranges considered, so our simulation with fixed conductivities overestimates the achieved temperature.
Based on our simulations, we estimate an upper bound maximum temperature of 200~K at the laser spot with a full-width-half-maximum of roughly 5~\textmu m (Fig. \ref{fig:thermcond}). 
Since these centers are stable up to room temperature and stand flash anneals at 1000$^\circ$C, the hypothesis of local annealing is thus inconsistent with our observations. 
}

\edit{ 
Our leading hypothesis is that the spectral shifts observed stem from DC Stark tuning arising from optically-induced charge variations, and that deactivation stems from ionization into the dark charged state of the G-center~\cite{udvarhelyi_identification_2021}.
Photoinduced spectral shifts have been previously reported in diamond color centers, and have been associated with trapped charges in the lattice~\cite{bassett_electrical_2011}.
It is well known that Si/SiO$_2$ interfaces contain a large density of hole traps~\cite{strand_intrinsic_2018} with an experimentally-measured saturation at densities in the order of $\sigma_h=1\times10^{13}$~cm$^{-2}$~\cite{afanasev_injection_1994}.
We hypothesize two flavors of this effect depending on initial conditions: 1) in the presence of existing charges, optical excitation will recombine them away, resulting a more neutral state, and 2) in the absence of charges, carrier generation may result in higher densities of trapped holes.
In both cases, the relative charge difference is the same, and thus in the following we estimate the effect that such charge variation can have on an artificial atom via DC Stark tuning.
Given the large difference in area between our single emitters and the waveguide surface, we can assume an infinitely large charged sheet, with an associated external electric field of $E=\frac{\sigma}{2\epsilon}$. 
$\sigma$ is the charge density, which for SiO$_2$ is in the order of $e\sigma_h$, with $e$ the elementary charge.
This estimate leads to electric fields in the order of $100$~MV/m.
Assuming DC Stark tuning rates in the order of $R=10$~GHz/MV m, in line with other color centers such as NV in diamond~\cite{bassett_electrical_2011}, we estimate maximum spectral shifts in the order of $\Delta \lambda_0=R \times E=1000$~GHz.
We note that our estimated maximum spectral shift will be significantly affected by our device geometry and oxide interface quality (likely different between the native oxide and thermally-grown oxide at the top and bottom of our waveguide), our estimate of Stark tuning rate, or other potentially more stringent limits such as defect ionization or charge recombination effects.
For high irradiation powers, one explanation is that ionization takes place, similarly to the previously reported photoionization of the NV color center in diamond~\cite{aslam_photo-induced_2013}.
}

\newpage
\subsection{\edit{Second-order correlation with a three-level atom model}}
\label{sec:g2_3ls}

\begin{figure*}[hbtp!]
  \centering
  \includegraphics[width=\textwidth]{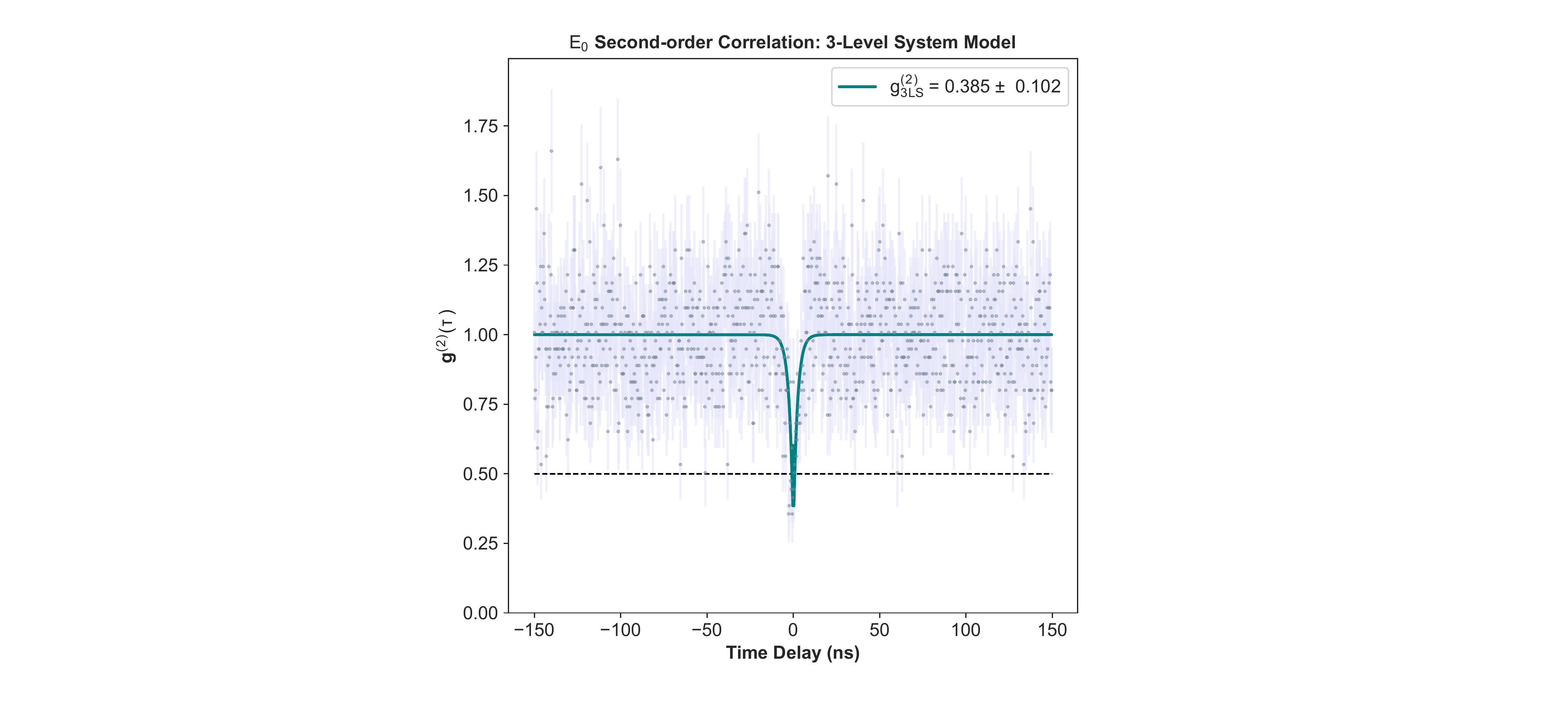}
\caption{\edit{\textbf{Second-order correlation fit with a three-level model.} The second-order correlation data, measured at 6~K under 10~$\mu$W continuous-wave 532~nm excitation power, corresponding to an estimated power density of $5.4$~kW$\cdot\text{cm}^{-2}$.}}
\label{fig:g2_3ls}
\end{figure*}

\edit{
The electronic level structure of the G-center defect has long been reported to consist of two singlet states and a metastable triplet state~\cite{lee_optical_1982, udvarhelyi_identification_2021}. 
The signature bunching effect in the second-order correlation, characteristic of a recombination process mediated by a three-level system, has been experimentally observed for G-centers in bulk SOI wafers~\cite{redjem_single_2020, hollenbach_engineering_2020}.
At the excitation powers used in our measurements, we would expect to observe slight bunching near the antibunching dip at a time delay of zero. 
However, this measurement was limited by noise resulting from low count emitter count rates due to poor mode matching between the waveguide and collection fiber. 
For completeness, we include in Figure~\ref{fig:g2_3ls} the second-order correlation data presented in the main text using a 3-level system fit: 
}
\edit{
\begin{equation}
    g^{(2)}_{3LS}(t) = a\left(1 - (1-b)e^{(-|t-t_{\text{shift}}|)/\tau_1)} + be^{(-|t-t_{\text{shift}}|/\tau_2))}\right)
\end{equation}
}

\appendix*

\end{document}